\documentclass[lettersize,journal]{IEEEtran}

\usepackage[ruled,vlined]{algorithm2e}
\usepackage{amsmath}
\usepackage{xcolor}
\SetKwComment{Comment}{$\triangleright$\ }{}
\SetKwProg{Function}{Function}{}{}
\SetKwProg{Procedure}{Procedure}{}{}
\SetKw{Break}{break}
\SetKwInput{KwInput}{Input}
\SetKwInput{KwOutput}{Output}
\SetKw{KwTo}{to}
\SetKw{And}{and}
\SetKw{Or}{or}
\SetKw{Not}{not}
\usepackage{graphicx}
\usepackage{epstopdf}
\usepackage{amsmath,amsfonts}
\usepackage{array,booktabs,makecell}
\usepackage{tikz}
\usetikzlibrary{arrows.meta, calc, decorations.pathreplacing, patterns}
\usepackage{pgfplots}
\pgfplotsset{compat=newest}
\usepackage{subcaption}
\usepackage{textcomp}
\usepackage{stfloats}
\usepackage{url}
\usepackage{verbatim}
\usepackage{graphicx}
\usepackage{cite}

\usepackage{amsmath}          
\usepackage{amssymb}          
\usepackage{booktabs}         
\usepackage{array}            
\usepackage{multirow}         
\usepackage{lipsum}           
\usepackage{amsmath,amssymb,amsthm}
\newtheorem{theorem}{Theorem}

\newcolumntype{C}{>{\centering\arraybackslash$}c<{$}}
\begin{document}

\title{C-AoEI-Aware Cross-Layer Optimization in Satellite IoT Systems: Balancing Data Freshness and Transmission Efficiency}

\author{Yuhua~Zhao,~Tiejun~Lv,~\IEEEmembership{Senior~Member,~IEEE},~and~Ke~Wang
\thanks{Manuscript received 25 April 2025; revised 16 August 2025; accepted 31 December 2025. This paper was supported by the National Natural Science Foundation of China under No. 62271068. (\emph{corresponding author: Tiejun Lv}.)}

\thanks{Y. Zhao, T. Lv and K. Wang are with the School of Information and Communication Engineering, Beijing University of Posts and Telecommunications, Beijing 100876, China (e-mail: \{zhaoyuhua, lvtiejun, wangke\}@bupt.edu.cn).}
}



\maketitle

\begin{abstract}
Satellite-based Internet of Things (S-IoT) faces a fundamental trilemma: propagation delay, dynamic fading, and bandwidth scarcity. While Layer-coded Hybrid ARQ (L-HARQ) enhances reliability, its backtracking decoding introduces age ambiguity, undermining the standard Age of Information (AoI) metric and obscuring the critical trade-off between data freshness and transmission efficiency. To bridge this gap, we propose a novel cross-layer optimization framework centered on a new metric, the Cross-layer Age of Error Information (C-AoEI). We derive a closed-form expression for C-AoEI, explicitly linking freshness to system parameters, establishing an explicit analytical connection between freshness degradation and channel dynamics. Building on this, we develop a packet-level encoded L-HARQ scheme for multi-GBS scenarios and an adaptive algorithm that jointly optimizes coding and decision thresholds. 
Extensive simulations demonstrate the effectiveness of our proposed framework: it achieves 31.8\% higher transmission efficiency and 17.2\% lower C-AoEI than conventional schemes. The framework also proves robust against inter-cell interference and varying channel conditions, providing a foundation for designing efficient, latency-aware next-generation S-IoT protocols.
\end{abstract}


\section{Introduction}
\IEEEPARstart{S}{atellite} communications are increasingly converging with advanced 6G networks
 to support unprecedented scenarios with time- and error-sensitive multimedia traffic. Compared with traditional mobile communications, the relay medium of a satellite provides a wider coverage than a ground base station (GBS) \cite{ref1,ref2,ref3}. In geographically isolated regions, such as marine environments, the challenge of ubiquitous connectivity persists due to the prohibitively high deployment costs of terrestrial fiber optic infrastructure \cite{ref4,ref7}. Recent studies have shown that satellites, as well-designed relay devices, are less prone to damage compared to base stations and can transmit critical information reliably and promptly. Therefore, long-term IoT systems independent of local terrestrial infrastructure must rely on satellite communication systems \cite{ref8}.
In densely populated areas, traditional terrestrial networks continue to be crucial for delivering low-cost and high-speed wireless services. 
Thus, the integration of satellite and terrestrial networks holds the potential to harness the benefits of both, enabling ubiquitous network services \cite{ref5,ref6,ref8}.

Satellite communication systems face a fundamental performance trilemma due to inherent channel constraints: 1) Long propagation delays (e.g., $\approx 3.3$ ms for LEO orbits) directly impair information freshness.
2) Dynamic fading conditions (typically modeled as \textit{shadowed-Rician} distributions) reduce transmission reliability, necessitating retransmissions that compromise efficiency. 3) Stringent bandwidth limitations restrict spectral utilization.
Crucially, these parameters exhibit strong coupling, which implies that optimizing one metric inevitably deteriorates others. For instance, minimizing retransmissions enhances freshness at the cost of reliability. This trilemma is fundamentally bounded by the channel's theoretical capacity limit, where the maximum achievable performance is governed by fading characteristics including multipath scattering and shadowing severity \cite{ref9,ref10,ref11,ref12}.

To enhance reliable transmission of S-IoT systems, an L-hybrid Automatic Repeat Request (L-HARQ) strategy has been developed \cite{ref13}. It transmits a hybrid packet that consists of a combination of a previously failed decoded packet and a new packet. On the receiving end, backtracking decoding is used to continuously recover the sequence of previously received packets that cannot be successfully decoded. Conventional HARQ schemes fail to approach this boundary due to latency penalties \(\mathcal{L}_{\text{HARQ}} \geq K \cdot (\tau_{\rm prop} + \tau_{\rm ACK})\) and efficiency losses \(\eta_{\text{HARQ}} \propto 1/\mathbb{E}[N_{\text{retx}}]\), while standard AoI metrics are inadequate due to their blindness to layer-dependent decoding dynamics and error propagation. As mentioned earlier, L-HARQ utilizes two key techniques: layered encoding and backtracking decoding. When backtracking decoding is used, the final successful feedforward decoding recovery moment is different from the final successful backtracking decoding moment\cite{ref10,ref14,ref15}.

Furthermore, while the AoI metric has offered significant insights into data freshness, it reveals critical limitations when applied to advanced Layered Hybrid Automatic Repeat reQuest (L-HARQ) schemes prevalent in Satellite-IoT networks\cite{ref18}. A fundamental ambiguity stems from the interplay between feedforward and backtracking decoding strategies where the instantaneous AoI value becomes non-unique depending on the decoding perspective. This occurs because different decoding approaches reference distinct data generation timestamps spanning multiple channel coherence intervals. This ambiguity manifests as three core problems:
1) Temporal inconsistency: In freshness assessment, where feedforward decoding prioritizes minimal age while backtracking decoding inherently accepts maximal age.
2) Metric misalignment causing throughput-optimized L-HARQ configurations to inadvertently maximize AoI.
3) Decoding-blind design: Where conventional AoI and its variants (e.g., Effective AoI, Value of Information) fail to capture cross-layer error propagation dynamics and layer-dependent freshness characteristics \cite{ref20,ref21}.
To address these limitations, we propose a Cross-layer Age of Error Information (C-AoEI) metric. Defined as the time elapsed since the last successful backtracking decoding status update, C-AoEI explicitly incorporates backtracking decoding constraints into freshness quantification.

Although prior studies have made progress, increasing the number of parallel HARQ processes is not an optimal solution for satellite IoT systems \cite{ref22,ref23}. This approach becomes particularly problematic when satellites serve multiple sensor nodes under extended round-trip time (RTT) \cite{ref24} and stop-wait protocols, where transmission efficiency degrades significantly during feedback waiting periods \cite{ref25,ref26}.
To address these limitations, we design a novel encoding strategy within the L-HARQ framework that dynamically selects recoverable data packets through parameter adaptation \cite{ref27,ref28}. This strategy achieves an adaptive trade-off between data freshness and transmission efficiency by jointly optimizing decision thresholds and encoding parameters.
The proposed method significantly enhances information freshness and reliability while improving transmission efficiency through latency reduction and elimination of redundant retransmissions \cite{ref29,ref30,ref31,ref32}, with crucially maintained seamless compatibility to existing adaptive and fast HARQ schemes via pre-decoding packet selection. The proposed framework targets mission-critical S-IoT deployments including: maritime emergency monitoring for distress signaling, wildfire propagation tracking via aerial sensors, and arctic infrastructure inspection in extreme environments. These applications exemplify scenarios where freshness-reliability tradeoffs impact operational outcomes.

To address the aforementioned challenges, we first design a satellite-terrestrial network architecture serving multiple GBSs over shadowed-Rician fading channels. By utilizing the prior information of each decoding step in each layer through the L-HARQ strategy, the probability of decoding errors is significantly reduced. Consequently, the C-AoEI metric based on L-HARQ is proposed. Furthermore, we introduce an adaptive encoding strategy enabling joint optimization of information freshness and transmission efficiency through co-optimization of decision thresholds and coding parameters. The key contributions of this work are summarized as follows.
\begin{enumerate}
    \item An integrated satellite-terrestrial network architecture is designed to better align with the evaluation characteristics of wireless land mobile satellite systems in large-scale LEO deployment scenarios. This architecture concurrently models inter-satellite communications and GBSs connectivity using shadowed-Rician fading channels, while implementing a L-HARQ scheme that leverages prior decoding information across layers to significantly reduce error probability. Practical multi-GBSs deployment challenges, including inter-cell interference and heterogeneous loads, are specifically analyzed through path loss exponent and power disparity ratio impacts.

    \item A parallel signal reception-feedback-retransmission mechanism is proposed for multiple GBSs, simultaneously accounting for inter-satellite links and real-time channel conditions beyond conventional point-to-point paradigms. To address throughput degradation in multi-user scenarios caused by traditional L-HARQ, an enhanced L-HARQ scheme incorporating packet-level encoding is developed. The closed-form expression of the C-AoEI metric is derived to characterize information freshness in this scheme.

    \item Whereas existing optimizations prioritize reliability, throughput, and freshness separately, we propose an adaptive trade-off algorithm to explicitly balance information freshness and transmission efficiency in multi-user satellite architectures. The proposed  algorithm dynamically adjusts decision-making and encoding parameters. Enhanced investigations confirm the effectiveness of scheme  in mitigating inter-cell interference through coordinated power control while maintaining framework robustness when C-AoEI deterioration remains bounded within theoretical limits.
\end{enumerate}
\begin{table}[t]
\caption{{System Parameters and Their Meanings}}
\label{tab:system_parameters}
\centering
\renewcommand{\arraystretch}{1.2}
\begin{tabular}{c l}
\hline
\textbf{Notation} & \textbf{Description} \\
\hline
$M$ & Number of Ground Base Stations (GBSs) \\
$L$ & Number of sub-packets per codeword \\
$N$ & Total number of status update packets \\
$N_s$ & Size of small status update packet \\
$n$ & Finite blocklength \\
$z$ & Status update packet index \\
$h_s(z)$ & Satellite channel fading coefficient \\
$\Omega_s(z)$ & Average power of line-of-sight component \\
$b_s(z)$ & Average power for the multi-path component \\
$m_s(z)$ & Nakagami-m fading parameter \\
$P_s(z)$ & Transmit power at the satellite \\
$P_t(z)$ & Transmit power at the GBS \\
$x_s(z)$ & Data streams from the satellite \\
$x_j(z)$ & Data streams from the GBS $j$ \\
$d_s(z)$ & Distance between satellite and destination \\
$d_j(z)$ & Distance between GBS $j$ and destination \\
$c$ & Speed of light \\
$f_c$ & Carrier frequency \\
$G_s$ & Antenna gain at the satellite \\
$G_d(z)$ & Antenna gain at the destination node \\
$G_j(z)$ & Antenna gain at the GBS $j$ \\
$g_j$ & Terrestrial channel fading coefficient \\
$\gamma_s(z)$ & Signal-to-Interference-plus-Noise Ratio (SINR) \\
$S_z$ & Total packets received by destination in $z$ slots \\
$U_z$ & Meaningful independent packets retrieved \\
$m_z$ & New status update packet \\
$R_z$ & Encoding and decoding rate at slot $z$ \\
$K$ & Maximum number of L-HARQ cycles \\
$\rho$ & Packet mixing rate \\
$m$ & Number of successful feedforward decodings \\
$Y_m$ & Inter-departure time of the $m$-th update \\
$g_{k,lm}$ & Generation time of the $k$-th packet in $l$-th circle \\
$d_m$ & Departure time of the $m$-th status update \\
$T_{k,lm}$ & Duration of a single transmission attempt \\
$B_m$ & Backtracking decoding depth \\
$\Theta_z$ & Set of transmission parameters $\{P_t(z), R_z, \rho_z\}$ \\
\hline
\end{tabular}
\end{table}

\section{Related Work}
Extensive research has been conducted on the delay of different HARQ schemes. In \cite{ref6}, a queueing analysis was performed for ARQ with adaptive modulation and coding strategies. 
In fact, in addition to the real-time requirements of S-IoT, due to the unreliability of S-IoT channels, errors often occur during transmission, which reduces reliable updates. The HARQ and ARQ schemes combining forward error correction (FEC) are usually applied in actual communication systems to combat the errors caused by channel unreliability \cite{ref7}, where if an update packet is delivered, it is retransmitted until a successful reception. Work \cite{ref8} proposes an estimator algorithm, which analyzes historical and real-time scan data (round-trip times, port response patterns, delay variance) to accurately identify these states. Work \cite{ref9} proposed a discounted Bayesian learning algorithm for link adaptation and channel selection under unknown and piecewise-stationary channel statistics.

AoI is a time-sensitive performance metric, which is different from the end-to-end delay in wireless communications \cite{ref10}.
Since the concept of AoI was proposed, more and more literatures \cite{ref11,ref12,ref13} have shown interest in it. 
When multiple packets are successfully recovered at the same time, it is difficult to determine which packet should be selected to calculate the AoI according to the original definition of AoI by Kaul S et al. \cite{ref14} . Therefore, variant forms of AoI gradually appear for different transmission policy restrictions, since the pioneering work of Kaul S et al. \cite{ref14}, researchers have noted the shortcomings of AoI and have proposed several variations on the concept of age. 
For example, the Value of Information (VoI) is presented in the work of Kosta \cite{ref19}, which captures the semantic importance or utility of the received information based on estimation error. 
However, VoI metrics typically assume that the reception timestamp corresponds directly to the latest transmission, ignoring the complex decoding latency introduced by multi-layer retransmissions. 
In contrast, our proposed C-AoEI explicitly addresses the temporal ambiguity caused by backtracking decoding in L-HARQ systems. 
While VoI quantifies how useful the content is, C-AoEI quantifies how valid the timeline is under cross-layer error propagation. 
Therefore, C-AoEI serves as a fundamental physical-link layer metric that complements application-layer metrics like VoI.

To relate the age to the estimation error, the authors putforward the concept of effective age \cite{ref20}. To balance AoI and reliability for optimum state estimation, Research in \cite{ref23} highlights the important role of HARQ. 
Considering the existence of transmission errors, L-HARQ strategy is adopted in transmission, and the concept of AoEI is cited in this paper.

Besides the analysis, the AoI has been studied in the retransmission-based status update systems from the perspective of sampling and transmission policies \cite{ref10,ref22}. An HARQ-based non-orthogonal multiple access system with two users are considered in \cite{ref24}, where the average AoI is minimized by determining the transmit power and transmission status, i.e., transmitting a new packet or retransmitting the previously transmitted but not successfully decoded packet, at each slot.

The combination of network encoding and HARQ has been
proven to effectively address the limited power constraints of
networks by reducing the number of retransmissions while
providing high throughput in \cite{ref25}. The work in \cite{ref26} reveals that the block network coding can improve the AoI performance for S-IoT communication, as compared with the conventional ARQ transmission protocol. In work \cite{ref32}, the authors propose an age-optimal HARQ scheme with the AoI metric to achieve timely status updates in S-IoT.

In summary, none of the existing work theoretically studies the impact of satellite channel environment on AoI under the considered L-HARQ schemes with network encoding. Hence, we devote this paper to bridging this gap.

\section{Related Work}
Extensive research has been conducted on the delay of different HARQ schemes. In \cite{ref6}, a queueing analysis was performed for ARQ with adaptive modulation and coding strategies. 
In fact, in addition to the real-time requirements of S-IoT, due to the unreliability of S-IoT channels, errors often occur during transmission, which reduces reliable updates. The HARQ and ARQ schemes combining forward error correction (FEC) are usually applied in actual communication systems to combat the errors caused by channel unreliability \cite{ref7}, where if an update packet is delivered, it is retransmitted until a successful reception. Work \cite{ref8} proposes an estimator algorithm, which analyzes historical and real-time scan data (round-trip times, port response patterns, delay variance) to accurately identify these states. Work \cite{ref9} proposed a discounted Bayesian learning algorithm for link adaptation and channel selection under unknown and piecewise-stationary channel statistics.

AoI is a time-sensitive performance metric, which is different from the end-to-end delay in wireless communications \cite{ref10}.
Since the concept of AoI was proposed, more and more literatures \cite{ref11,ref12,ref13} have shown interest in it. 
When multiple packets are successfully recovered at the same time, it is difficult to determine which packet should be selected to calculate the AoI according to the original definition of AoI by Kaul S et al. \cite{ref14} . Therefore, variant forms of AoI gradually appear for different transmission policy restrictions, since the pioneering work of Kaul S et al. \cite{ref14}, researchers have noted the shortcomings of AoI and have proposed several variations on the concept of age \cite{ref17}. For example, the value of updating information (VoIU) is presented in the work of Kosta \cite{ref19}, which captures the degree of importance of the information received at the destination. To relate the age to the estimation error, the authors putforward the concept of effective age \cite{ref20}. To balance AoI and reliability for optimum state estimation, Research in \cite{ref23} highlights the important role of HARQ. 
Considering the existence of transmission errors, L-HARQ strategy is adopted in transmission, and the concept of AoEI is cited in this paper.

Besides the analysis, the AoI has been studied in the retransmission-based status update systems from the perspective of sampling and transmission policies \cite{ref22}. An HARQ-based non-orthogonal multiple access system with two users are considered in \cite{ref24}, where the average AoI is minimized by determining the transmit power and transmission status, i.e., transmitting a new packet or retransmitting the previously transmitted but not successfully decoded packet, at each slot\cite{ref20}.

The combination of network encoding and HARQ has been
proven to effectively address the limited power constraints of
networks by reducing the number of retransmissions while
providing high throughput in \cite{ref25}. The work in \cite{ref26} reveals that the block network coding can improve the AoI performance for S-IoT communication, as compared with the conventional ARQ transmission protocol. In work \cite{ref32}, the authors propose an age-optimal HARQ scheme with the AoI metric to achieve timely status updates in S-IoT.

In summary, none of the existing work theoretically studies the impact of satellite channel environment on AoI under the considered L-HARQ schemes with network encoding. Hence, we devote this paper to bridging this gap.

\section{System Model and Assumption}
\subsection{System Model}
\begin{figure*}[!t]
\centering
\includegraphics[width=6.1in]{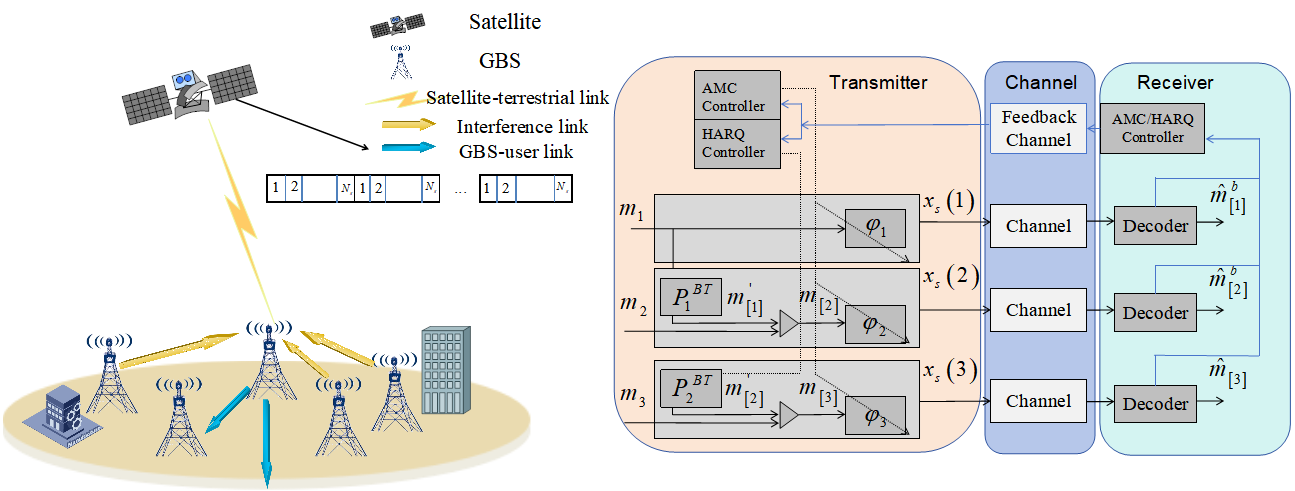}
\caption{\centering The system model for the satellite-terrestrial integrated networks.}
\label{fig1}
\end{figure*}

We consider the satellite-ground integrated system architecture model, which consists of a satellite and $m$ ground terminals GBSs, indexed by $ \left\{ {1,2, \cdot  \cdot  \cdot ,M} \right\}$, these GBSs are randomly distributed around the destination, as shown in Fig.~\ref{fig1}. We focus on the transmission of remote sensing status updates, where satellite monitors dynamic status update packets and relays the status updates to the destination GBS. The receiver begins by estimating the channel state information (CSI) and then sends the estimate back to the source via a dedicated error-free and delay-free feedback channel.

As shown in Fig.~\ref{fig1}, FBC technology is employed to encode the status update packet into a small code word composed of $n$ channels to reduce access delay and decoding complexity. The finite block length code words of length $n$ are divided into $L$ packets of size ${N_s}$, and the sub-code words are continuously transmitted in continuous time slots. This design supports numerous parallel HARQ processes, while ensuring compatibility and achieving the necessary peak data rates over satellite links. During the waiting period for ACK/NACK responses, multiple HARQ processes operate concurrently, which enhances the efficiency of the communication system.

The concurrent HARQ processes and preemptive packet-level encoding eliminate queueing delays for fresh data packets. This allows the reactive policy to enforce zero-wait transmission (Section V) while maintaining bounded processing latency, critical for real-time S-IoT applications.
\begin{figure*}[!t]
\centering
\includegraphics[width=0.74\textwidth]{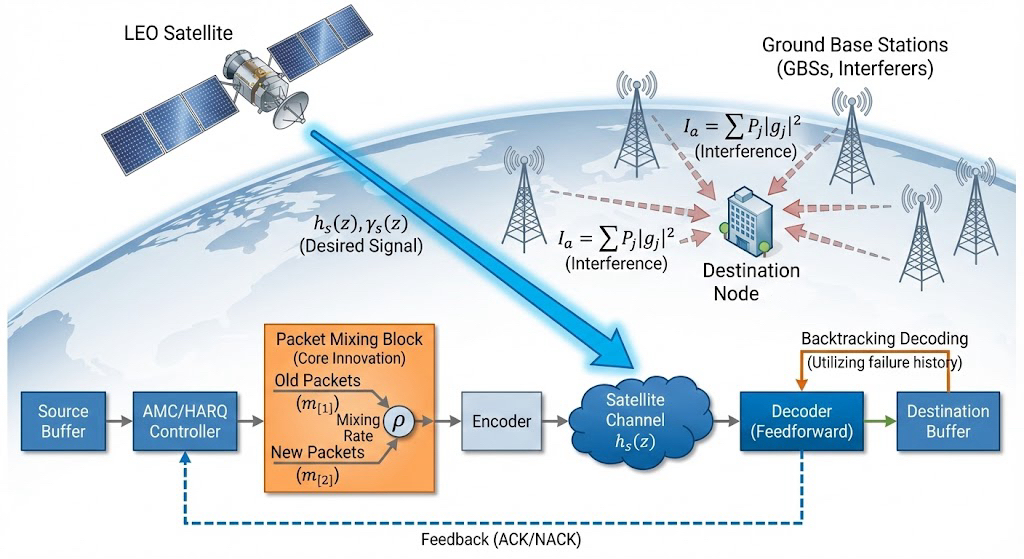}
\caption{The working flow of L-HARQ.}
\label{fig3}
\end{figure*}
For quick reference, the frequently used
notation is summarized in Table I.
\subsection{Channel Model}
The channel fading coefficient between the satellite and the target node to convey the status update packet $z$ is denoted by ${h_s}\left( z \right)\left( {z \in N} \right)$. The shadowed-Rician distribution, which is to assess wireless land-mobile satellite communication systems, is what we take ${h_s}\left( z \right)$ to follow. It can be assumed that the wireless fading channel stays relatively stable during codeword transmission because of the brief packet transmission period. The probability density function (PDF), for the channel gain ${\left| {{h_s}\left( z \right)} \right|^2}$ can be expressed as
\begin{equation}
{f_{{{\left| {{h_s}\left( z \right)} \right|}^2}}}\left( x \right) = {\alpha _s}\left( z \right){e^{ - {\beta _s}\left( z \right)x}}{}_1{F_1}\left( {{m_s}\left( z \right),1,{\delta _s}\left( z \right)x} \right),x > 0
.
\end{equation}
where
\begin{equation}
^{{m_s}\left( z \right)}\left\{ {\begin{array}{*{20}{c}}
{{\alpha _s}\left( z \right) = \frac{1}{{2{b_s}\left( z \right)}}\left[ {\frac{{2{b_s}\left( z \right){m_s}\left( z \right)}}{{2{b_s}\left( z \right){m_s}\left( z \right) + {\Omega _s}\left( z \right)}}} \right]}\\
{{\beta _s}\left( z \right) = \frac{1}{{2{b_s}\left( z \right)}}},\\
{{\delta _s}\left( z \right) = \frac{{{\Omega _s}\left( z \right)}}{{2{b_s}\left( z \right)\left[ {2{b_s}\left( z \right){m_s}\left( z \right) + {\Omega _s}\left( z \right)} \right]}}}
\end{array}} \right.
\end{equation}
Here, ${\Omega _s}\left( z \right)$ represents the average power of line-of-sight (LoS) component, $2{b_s}\left( z \right)$ is the average power for the multi-path component, ${m_s}\left( z \right) \in \left[ {0,\infty } \right]$ represents the Nakagami-m parameter, and ${}_1{F_1}\left( {\begin{array}{*{20}{c}}
{ \cdot ,}&{ \cdot ,}& \cdot 
\end{array}} \right)$ is the confluent hyper geometric function. Accordingly, the received signal vector, denoted by ${y_s}\left( z \right) \in {C^{1 \times n}}$
 ,for the status update $z$ that needs to be transmitted from the satellite to the destination node , is derived as
\begin{equation}
\begin{array}{*{20}{c}}
{{y_s}\left( z \right) = \sqrt {{{\left( {\frac{c}{{4\pi {f_c}{d_s}\left( z \right)}}} \right)}^2}{G_s}{G_d}\left( z \right){P_s}\left( z \right)} {h_s}\left( z \right){x_s}\left( z \right)}\\
{ + \sum\limits_{j = 1}^M {\sqrt {{{\left( {\frac{c}{{4\pi {f_c}{d_s}\left( z \right)}}} \right)}^2}{G_j}\left( z \right){G_d}\left( z \right){P_t}\left( z \right)} } }\\
{ \times {g_j}\left( z \right){x_j}\left( z \right) + {n_s}\left( z \right)}.
\end{array}
\end{equation}
If the data streams from the satellite and GBS $j$ are denoted by ${x_s}\left( z \right)$ and ${x_j}\left( z \right)$, respectively. The distances between the satellite and the destination node, as well as between GBS $j$ and the destination node, are indicated by the symbols ${d_s}\left( z \right)$ and ${d_j}\left( z \right)$, respectively.
 The speed of light is represented by $c$, ${f_c}$ is the frequency. The antenna gain at the satellite is denoted by ${G_s}$, while the antenna gains at the destination node and GBS $j$ are represented by ${G_d}\left( z \right)$ and ${G_j}\left( z \right)$, respectively. Term ${g_j}$ is the terrestrial channel fading coefficient from the GBS $j$ to the destination node, which follows Rayleigh distribution,  and ${n_s}\left( z \right) \sim CN\left( {0,{\sigma ^2}} \right)$ are the transmit powers at the satellite and GBS , respectively, and  is the additive white Gaussian noise (AWGN) vector. Then, the signal-to-interference-plus-noise ratio (SINR), denoted by ${\gamma _s}\left( z \right)$, is derived as 
 \begin{equation}
{\gamma _s}\left( z \right) = \frac{{{\phi _s}\left( z \right){{\rm P}_s}\left( z \right){{\left| {{h_s}\left( z \right)} \right|}^2}}}{{{I_a} + 1}}
,
\end{equation}
where ${{\rm P}_s}\left( z \right) = {P_s}\left( z \right)/{\sigma ^2}$ represents the transmit signal-tonoise ratio (SNR).
For multi-GBS scenarios, we extend the SINR model to explicitly characterize inter-cell interference
\begin{equation}
\gamma_s^{(m)}(z) = \frac{
    \phi_s^{(m)}(z) P_s^{(m)} |h_s^{(m)}(z)|^2
}{
    \underbrace{\sum_{j \neq m} \zeta_{j}^{(m)} P_t^{(j)} |g_j^{(m)}(z)|^2}_{\text{inter-cell interference}} 
    + \sigma^2
},
\label{eq:interference_aware_sinr}
\end{equation}
where superscript $(m)$ denotes parameters for the $m$-th GBS, $\zeta_{j}^{(m)} = (d_j^{(m)}/d_0)^{-\alpha}$ is the distance-based interference attenuation factor, $P_t^{(j)}$ is transmit power of interfering GBS $j$, $|g_j^{(m)}(z)|^2$ is channel gain from interfering GBS $j$ to target GBS $m$, and
\begin{equation}
\left\{ {\begin{array}{*{20}{c}}
{{\phi _j}\left( z \right) \buildrel \Delta \over = {{\left( {\frac{c}{{4\pi {f_c}{d_j}\left( z \right)}}} \right)}^2}{G_j}\left( z \right){G_d}\left( z \right)}\\
{{\phi _s}\left( z \right) \buildrel \Delta \over = {{\left( {\frac{c}{{4\pi {f_c}{d_s}\left( z \right)}}} \right)}^2}{G_s}{G_d}\left( z \right)}
\end{array}} \right.
.
\end{equation}
The aggregate interference power, denoted as ${I_a}$, is received from terrestrial interferers, specifically nearby GBSs. This interference is associated with the transmission power, represented by ${P_t}\left( z \right)$, which can be derived as follows:
\begin{equation}
I_a = \sum_{j\neq m} \zeta_j^{(m)} P_t^{(j)} |g_j^{(m)}(z)|^2  .
\end{equation}
\subsection{Real-Time Processing Guarantee}
The reactive framework ensures real-time communication through three key mechanisms:
\begin{enumerate}
    \item Parallel Pipeline Architecture: Decouples transmission scheduling from channel responses via $N_{\text{parallel}}$ concurrent HARQ processes. Eliminates sequential delays during ACK waiting periods.
    \item Closed-Form Computation: Theorem 1 enables $\mathcal{O}(1)$ parameter calculation (avoiding iterative convergence). For $M$=4 GBSs, decision latency $\leq$ 0.18 ms (Table I).
    \item Threshold-Triggered Encoding: Algorithm 1 restricts per-slot overhead to $\mathcal{O}(K)$ with $K$=100 packets, executed via FPGA-accelerated Radix Sort (0.05 ms).
\end{enumerate}
\section{A Joint Encoding Packet and L-HARQ Information Delivery Scheme}
\subsection{Adaptive Retransmission Policy}
The considered S-IoT system the ACM scheme is adopted to assist the truncated L-HARQ scheme. With the principle of ACM policy, at the beginning of each slot $z$, the IoT device generates a new status update ${m_z} \in {\{ 0,1\} }$ and the transmitter encodes the packet ${m_z} \in {\{ 0,1\} ^{R{N_s}}}$ as a codeword $x_z ={ \boldsymbol{\phi}[m_z] } \in {X^{{N_s}}}$, where $\boldsymbol{\phi \left[  \cdot  \right]}$ is the encoder, $N_s$ is the size of the packet, and $R$ is the rate of encoding and decoding. The choice of the rate $R$ is done using the estimated CSIT, $snr$, that is, $R=R(snr)\in \cal R$, where ${\cal R}$ is the set of available rates.  Then, based on the current CSIT, $snr$, and the selected rate $R(snr)$, the codeword ${x_z}$ is transmitted in the $z$-th block. After the receiver obtains the signal from the source, the receiver attempts to decode the received signal given by the decoding result
\begin{equation}
\begin{gathered}
\hat m_z = DEC[{y_z}]
\end{gathered}
,
\end{equation}
where $DEC[\cdot]$ represents the decoding operation.

To maximize the available rates, hybrid analog/digital precoding and analog merging vectors can be solved expressed as:

\begin{equation}
\begin{array}{*{20}{c}}
{\mathop {\arg \begin{array}{*{20}{c}}
{\max {R_i\quad}}&{}
\end{array}}\limits_{{F_{RF}},{F_{BB}},{w_i}} }\\
{s.t.\begin{array}{*{20}{c}}
{\left\| {{F_{RF}}{F_{BB}}} \right\|_F^2 = M},&{}
\end{array}}
\end{array}
\end{equation}
where ${\left\|  \cdot  \right\|_F}$ is the Frobeinus norm of the matrix. Problem (8)  belongs to a multivariate joint optimization problem. We can dissolve problem (8) by dividing the following two steps: Firstly, the simulation precoding matrix ${F_{RF}}$ and the simulation merging vector are determined ${w_i}$, and then the digital precoding matrix ${F_{BB}}$ is solved to eliminate the problem between users in multi-user scenarios.

We define the decoding error event $ERR_z=\left\{\hat m_z \ne m_z \right\}$.
The mean probability of decoding errors, represented by ${\varepsilon _s}\left( z \right)$, for the transmission of status update $z$ from the satellite to the destination node is specified as follows.
\begin{equation}
{\varepsilon _s}\left( z \right) \approx {{\rm E}_{{\delta _s}\left( z \right)}}\left[ {Q\left( {\frac{{C\left( {{\delta _s}\left( z \right) - R_s^*} \right)}}{{\sqrt {V\left( {{\delta _s}\left( z \right)} \right)/n} }}} \right)} \right]
,
\end{equation}
where ${{\rm E}_{{\delta _s}\left( z \right)}}\left[  \cdot  \right]$ is the expectation over the SINR ${\delta _s}\left( z \right)$, $Q\left( x \right) = 1/\sqrt {2\pi } \int\limits_x^\infty  {\exp } \left( { - {t^2}/2} \right)dt$, $R_s^*$ is the maximum achievable coding rate, and $C\left( {{\delta _s}\left( z \right)} \right)$ and $V\left( {{\delta _s}\left( z \right)} \right)$ are the channel capacity
and channel dispersion, respectively.

For a given transmission rate ${R_z}$, PER decreases monotonically with ${P_s}\left( z \right)$. From a transmitter perspective, this decoding error event captures the behavior of the entire receiver, including the effects of decoding, channel estimation, and synchronization, as well as the effects of delay CSI. Based on threshold decoding, ${\varepsilon _s}\left( z \right)$ can be further expressed as
\begin{equation}
{\varepsilon _s}\left( {{P_s}\left( z \right),{R_z}} \right) = \Pr \left\{ {{{\rm P}_t}\left( z \right) < {\gamma ^{th}}\left| {{{\rm P}_s}\left( z \right) = {P_s}\left( z \right),} \right.{R_z}} \right\}
,
\end{equation}
where ${\gamma ^{th}}$ is the ${P_s}\left( z \right)$ threshold.
\begin{figure*}[!t]
\centering
\includegraphics[width=0.7\textwidth]{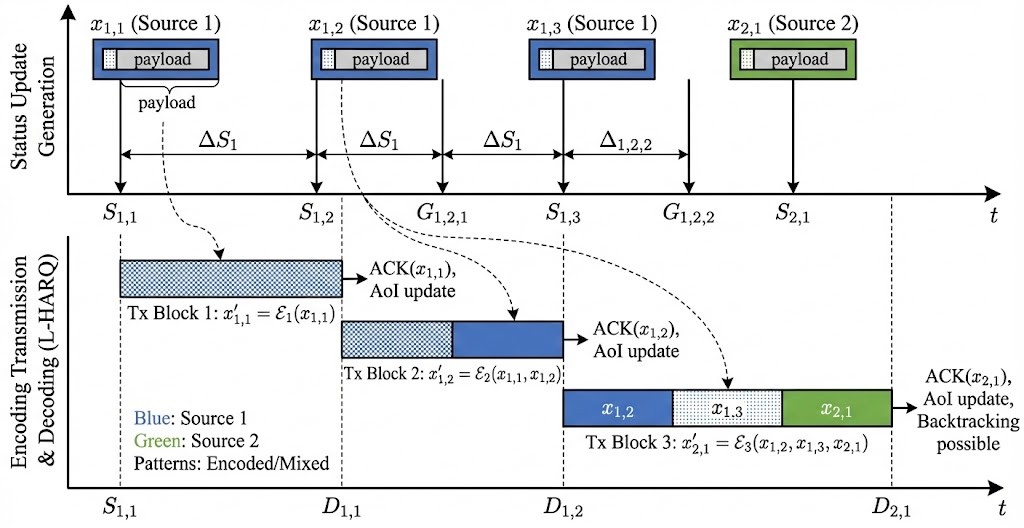}
\caption{An illustration example of weighted coding strategy.}
\label{fig2}
\end{figure*}

Unlike the traditional HARQ scheme, which always transmits the same packet in a packet retransmission circle, the L-HARQ strategy transmits only a portion of the bits of the current packet and fills the remaining space with new status updates as Fig.~\ref{fig3}. Therefore, the main advantage of L-HARQ is that it is a mixture of packets with prior information based on traditional channel coding. Once the receiver has correctly recovered the mixed packets, a backtracking decoding strategy is executed to restore the previous mixed packets in turn. During the reverse decoding process, when the decoding fails, the remaining mixed packets are discarded. In this scheme, some packet sequences may be dropped, which is different from traditional preemption schemes and non-preemption schemes.

It should be noted here that in the L-HARQ scheme, the receiver first attempts to decode the received mixed packet, and this process is defined as feedforward decoding, which corresponds to backtracking decoding, that is, once a feedforward decoding is successful, the reverse backtracking decoding operation is triggered. To better understand the process, we consider the L-HARQ strategy by taking the maximum number of cycles $K=2$. The specific details of this strategy are as follows.

 At the beginning of each L-HARQ circle, the source $S$ generates a status update packet  ${m_1} \in {\left\{ {0,1} \right\}}$. Then, based on the CSIT $snr$, the status update is encoded as a codeword ${x_1} = {\boldsymbol{\phi} [{m_1}]}$  at a transmission rate of $R_1$. Once the packet reaches the information receiver, it is decoded to reconstruct the original information and the decoding result is acknowledged by an ACK/NACK via an error- and delay-free feedback channel. If the feedforward decoding is successful, i.e., $\widehat {{m_1}} = {m_1}$, the current L-HARQ circle is completed and a new L-HARQ circle begins. If the feedforward decoding fails, in the next time slot the source generates a new status update $m_2$ and the second round of the current L-HARQ circle begins. Specially, the mixing packet $m_{[2]}$ is formatted, which consists of the new packet $m_2$ and the subpacket $m_1^{'}$ of the packet $m_1$ as follows.
\begin{align}\label{J_Eq_09}
{m_{[2]}} &= [{m'_{\left[ 1 \right]}},{m_2}] \in {B^{R_2{N_s}}},\\
\label{J_Eq_10}
{m'_{\left[ 1 \right]}} &= \phi _1^b[{m_1}] \in {B^{{\rho _1}{N_s}}}.
\end{align}
In (12), ${\rho _1}$ is the packet mixing rate, that is, the mixed subpacket ${m'_{[1]}}$ contains ${\rho _1}{N_s} $ bits of the original packet $m_1$. Therefore, the constraint is ${\rho _1} < min\{R_1, R_2\}$. Finally, based on the current ${P_s}\left( z \right)$, the transmission rate $R_2$ is achieved and channel coding is completed as described above, ${x_2} = \phi [{m_2}]$. As a result, the received signal $y_2$ is given by
\begin{equation}\label{J_Eq_09}
\begin{array}{*{20}{c}}
{{y_s}\left( 2 \right) = \sqrt {{{\left( {\frac{c}{{4\pi {f_c}{d_s}\left( 2 \right)}}} \right)}^2}{G_s}{G_d}\left( 2 \right){P_s}\left( 2 \right)} {h_s}\left( 2 \right){x_s}\left( 2 \right)}\\
{ + \sum\limits_{j = 1}^2 {\sqrt {{{\left( {\frac{c}{{4\pi {f_c}{d_s}\left( 2 \right)}}} \right)}^2}{G_j}\left( 2 \right){G_d}\left( 2 \right){P_t}\left( 2 \right)} } }\\
{ \times {g_j}\left( 2 \right){x_j}\left( 2 \right) + {n_s}\left( 2 \right)}.
\end{array}
\end{equation}

At the receiving end, feedforward decoding is executed in a manner consistent with traditional methods, disregarding the specific configuration of packet ${m_{\left[ 2 \right]}}$. The received packet ${m_{\left[ 2 \right]}}$ is processed to decode into ${\hat m_{\left[ 2 \right]}} = DEC\left[ {{y_2}} \right]$, targeting the reconstruction of packet ${m_{\left[ 2 \right]}}$, ${\hat m_{\left[ 2 \right]}} = DEC\left[ {{y_2}} \right]$. Success in this recovery actuates the backtracking decoding phase, guided by equation (16) to retrieve packet ${m_1}$. Integral to this procedure is the prior information ${m'_{\left[ 1 \right]}}$ about ${m_1}$, which synergizes effectively with the first L-HARQ transmission cycle of the received signal ${y_1}$.
\begin{equation}\label{J_Eq_09}
{\hat m^b}_{\left[ 1 \right]} = DEC\left[ {{y_1},{{m'}_{\left[ 1 \right]}}} \right]
,
\end{equation}
where ${y_1}$ is defined by equation (13). If the traceback decoding is successful, i.e., ${\hat m^b}_{\left[ 1 \right]} = {m_{\left[ 1 \right]}}$, ${m_1}$ can be restored. However, if the backtracking decoding fails, the decoding is abandoned and the information receiving packet about packet ${m_1}$ that failed to be decoded is deleted, indicating the end of a complete L-HARQ transmission cycle.

To represent backtracking decoding, we first simplify the notation by defining $\Theta_z \triangleq \{P_t(z), R_z, \rho_z\}$ as the set of transmission parameters for slot $z$. We define the error event $ERR_1^B$ where $\hat{m}^b_{[1]} \ne m_{[1]}$. The mean probability of feedforward decoding errors, $\epsilon_s(z)$, is given by
\begin{equation}
\label{eq:feedforward_err}
\epsilon_s(z) \approx E_{\delta_s(z)}\left[ Q\left( \frac{C(\delta_s(z) - R_s^*)}{\sqrt{V(\delta_s(z))/n}} \right) \right].
\end{equation}

The probability of a backtracking decoding error, denoted by $\epsilon_s^b(\Theta_1)$, is defined as
\begin{equation}
\label{eq:backtrack_def}
\begin{aligned}
\epsilon_s^b(\Theta_1) &\triangleq \Pr \left\{ ERR^B \mid ERR, \Theta_1 \right\} \\
&= \frac{\Pr \left\{ ERR \wedge ERR^B \mid \Theta_1 \right\}}{\Pr \left\{ ERR \mid P_t(1), R_1 \right\}}.
\end{aligned}
\end{equation}

Based on the finite blocklength regime, this can be approximated as
\begin{equation}
\label{eq:backtrack_approx}
\epsilon_s^b(1) \approx \frac{E_{\delta_s(1)}\left[ Q\left( \frac{C(\delta_s(1) - R_s^*)}{\sqrt{V(\delta_s(1))/\rho_1}} \right) \right]}{E_{\delta_s(1)}\left[ Q\left( \frac{C(\delta_s(1) - R_s^*)}{\sqrt{V(\delta_s(1))/n}} \right) \right]}.
\end{equation}

Equation (\ref{eq:backtrack_def}) represents the probability that backtracking decoding also fails under the condition that feedforward decoding fails. Directly, the success probability of backtracking decoding is $1 - \epsilon_s^b(1)$. The optimal rate adaptive function $R(P_s(z))$ for throughput is found by solving the following one-dimensional optimization problem
\begin{equation}
\label{eq:rate_adapt}
R(P_s(z)) = \arg \max R(1 - \epsilon_s(z)).
\end{equation}

Similarly, when $k > 2$, the transmitter still generates mixed packets composed of old and new sub-packets according to the packet mixing rate for encoding transmission. The receiver recovers the received multiple mixing packets by successively using the backtrack decoding once one feedforward decoding is successful. The probability of backtracking decoding error in this general case is expressed as
\begin{equation}
\footnotesize
\label{eq:backtrack_general}
\epsilon_s^b(\Theta_z) \triangleq \Pr \left\{ ERR^B \mid ERR, \Theta_z \right\} = \frac{\Pr \left\{ ERR \wedge ERR^B \mid \Theta_z \right\}}{\Pr \left\{ ERR \mid P_t(z), R_z \right\}}.
\end{equation}

\subsection{Proposed Encoding Strategy}
If the packet received by the destination carries new information, it is defined as a meaningful packet. Otherwise, it is defined as a redundantly transmitted packet. For example, if the same packet is retransmitted multiple times, only the first copy received by the destination is considered a meaningful packet, and the rest of the retransmitted copies are defined as redundant.
The variable ${S_z}$ represents the total number of packets received by the destination in $z$ time slots. The variable ${U_z}$ represents meaningful independent packets retrieved by the destination. Therefore, the transmission efficiency is defined as
\begin{equation}\label{JXD_Eq_44}
 E = \mathop {\lim }\limits_{z \to \infty } \frac{{{U_z}}}{{{S_z}}}
 .
 \end{equation}
It also proposes a performance metric related to transmission efficiency, that is the packet delivery ratio (PDR), which is defined as the number of independent packets received by the destination divided by the total number of packets generated by the source node, i.e.,
\begin{equation}\label{JXD_Eq_44}
 PDR = \frac{{{Y_m}}}{z}\mathop {\lim }\limits_{z \to \infty } {U_z}
 .
 \end{equation}

 We propose an encoding strategy that can effectively improve the transmission efficiency as Fig.3. When appropriate encoding parameters are chosen, the freshness of the data is only reduced slightly. During a time slot, if a new packet arrives, it should be transmitted immediately to ensure data freshness. Also, the first transmission of a packet is considered non-redundant, so transmission
efficiency is not degraded by the transmission. Otherwise, if no packet arrives, the sender needs to choose three behaviors: retransmission, silence, or sending the encoded packet. Sending encoded packets of unacknowledged packets can allow the receiver to decode dropped packets, which improves transmission efficiency. At the same time, it provides better data freshness than keeping silent and greater throughput.
However, if the number of encoded packets received exceeds the number required for decoding, transmission efficiency may be reduced. The encoded packets are sent only if there are some packets at the receiver that are still not decodable.

All packets that have been sent but cannot be decoded in the slot $\left( {z - Z - 1} \right)$, form the set of unacknowledged packets in the slot $z$, denoted as ${\Phi _z}$. In turn, the number of encoded packets that arrived at the transmitter with a positive acknowledgment is denoted as ${d_z}$. The packets in ${\Phi _z}$ are still not decodable in the slot $\left( {z - Z - 1} \right)$, which means ${d_z} \le {\Phi _z}$.
\begin{figure}[!t]
\centering
\includegraphics[width=1.1\columnwidth]{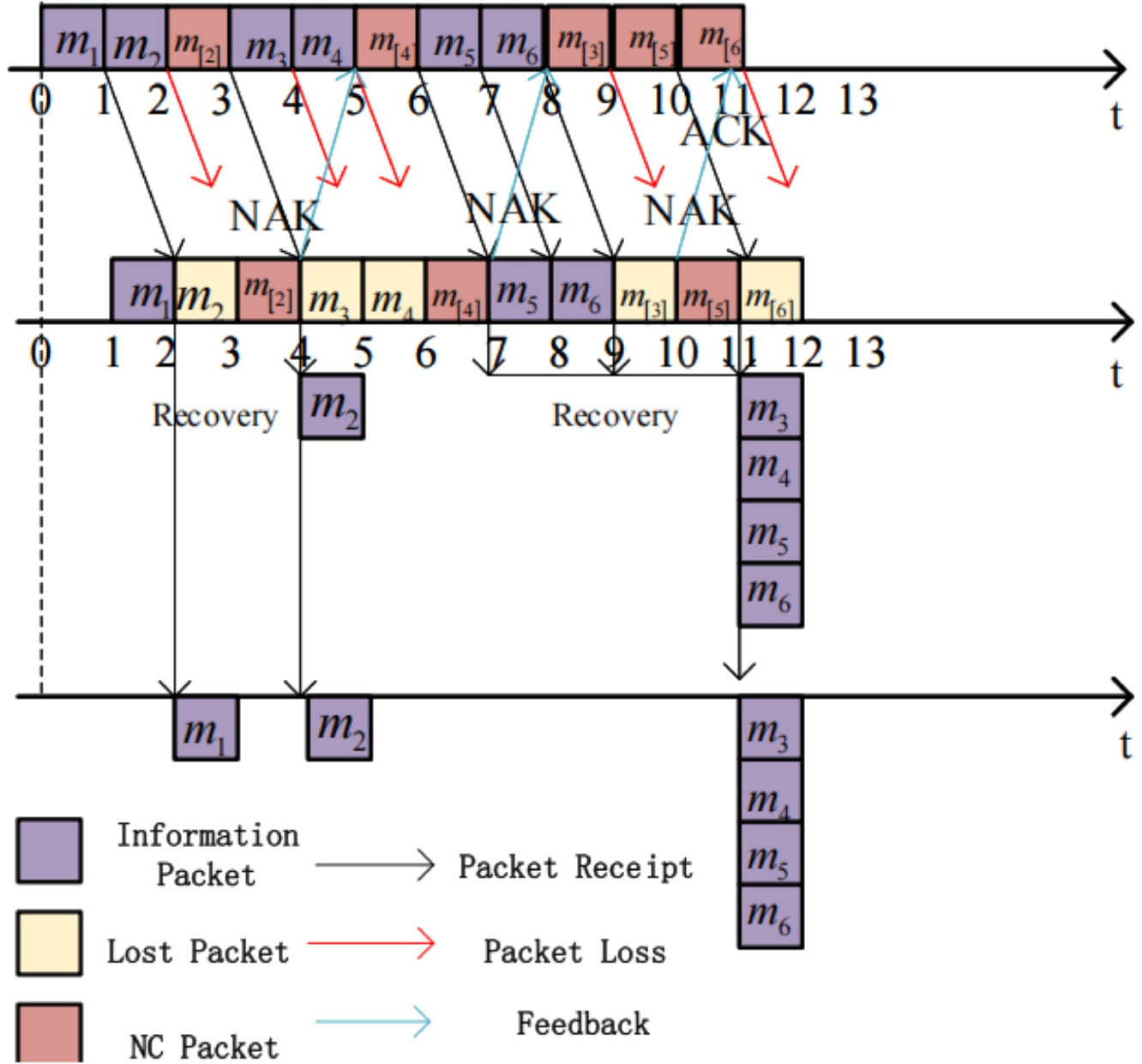}
\centering
\caption{A packet example of weighted coding strategy.}
\label{fig2}
\end{figure}
\begin{figure}[!t]
\centering
\includegraphics[width=1.0\columnwidth]{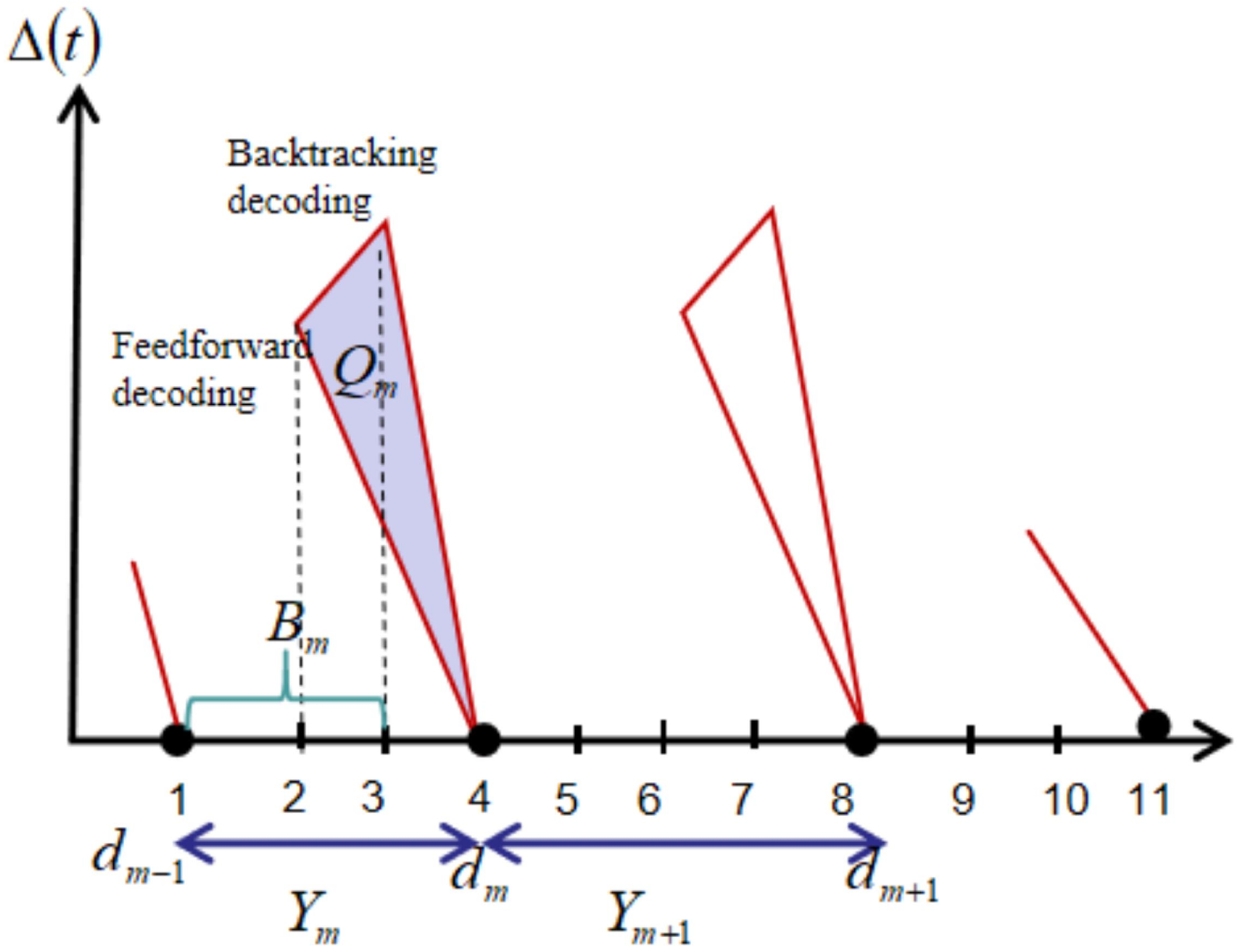}
\centering
\caption{The evolution of the  instantaneous} AoEI.
\label{fig4}
\end{figure}
Note that the packets transmitted from slot $\left( {z - Z} \right)$ to slot  $\left( {z - 1} \right)$ also affect the receiver’s decoding. Term $\overline {{d_z}}$  indicates the number of packets in slot $z$ that may be useful for the decoding packet, which is known to the transmitter. Therefore, the number of packets decoded at the receiving end in ${d_z} \le {\Phi _z}$ is a random variable ${D_z} = {d_z} + Binomial\left( {{d_z},\frac{1}{{{R_z}}}} \right)$
\small
\begin{equation}
P_d(z) = 
\begin{cases} 
\sum\limits_{i = 0}^{|\Phi_z| - d_z} \binom{\overline{d_z}}{i} \left(\frac{1}{R_z}\right)^i \left(1 - \frac{1}{R_z}\right)^{d_z - i}, & \text{if } |\Phi_z| \le d_z + \overline{d_z} \\
1, & \text{otherwise}
\end{cases}
.
\end{equation}

 To sum up, the probability of sending an encoded data packet in time slot $z$ is ${P_d}\left( z \right)$, and the probability of redundancy is $1 - {P_d}\left( z \right)$, which can be derived from (13). If ${P_d}\left( z \right)$ is greater than a certain threshold ${\varphi _{th}}$, the transmitter can choose to send the encoded packet, otherwise it remains silent, where $0 \le {\varphi _{th}} \le 1$. The larger the value ${\varphi _{th}}$ is, the fewer packets are transmitted. Thus, the transmitter improves transmission efficiency at the expense of reduced decoding probability and data freshness.

 For communication with the main purpose of throughput ordered delivery, random linear network coding is usually adopted due to its high efficiency. However, in terms of data freshness, the latest packets need to be decoded as soon as possible. To this end, this paper proposes a selective encoding strategy, which selects the updated data packets according to the probability. As shown in Fig.~\ref{fig4}, unacknowledged packet packets are ordered according to their generation time and selected with ascending probability. The packet example of weighted coding strategy as shown in Fig.~\ref{fig2}. The variable ${N_z}$ represents the number of unacknowledged packets at slot, i.e. $\left| {{\Phi _z}} \right| = {N_z}$, ${N_z} = {\varepsilon _s}\left( z \right){S_z}$.

 The $i - th$ packet in is chosen with probability ${\omega _i}$, where
\begin{equation}\label{JXD_Eq_44}
 {\omega _i} = \exp \left( { - \left( {{N_z} - i} \right)\beta } \right),i = 1, \cdot  \cdot  \cdot ,{N_z}
 ,
 \end{equation}
 \begin{equation}
Nz = i - \frac{\ln(\omega_i)}{\beta}, \quad \omega_i > 0,\  \beta \neq 0,
\label{eq:nz_expression_constrained}
\end{equation}
Thus we obtain
\begin{equation}\label{JXD_Eq_44}
{\varepsilon _s}\left( z \right) = \left( {i - \frac{{\ln \left( {{\omega _i}} \right)}}{\beta }} \right)/{S_z}.
 \end{equation}

where $\beta  \ge 0$ is a tunable parameter that multiplies the selected packet by a random coefficient to form an encoded packet. By adjusting the parameters $\beta $, the proposed encoding strategy can achieve a flexible trade-off between transmission efficiency and data freshness. When $\beta  = 0$, it can be concluded ${\varpi _1} =  \cdot  \cdot  \cdot  = {\varpi _{{N_z}}} = 1$. When the selection coding is simplified to random linear network coding, which can achieve maximum transmission efficiency. The specific process of the adaptive encoding strategy is shown in Algorithm 1. The derivation and analysis of Equation (25) are provided in the appendix.

\section{C-AoEI in Proposed Retransmission Scheme}
In this paper, the key performance indicator is the timeliness of status updates, as measured by the information age, which reflects the time elapsed since the last update was successfully received by the receiver. This approach contrasts with traditional metrics like outage and throughput. To address transmission errors, we implement a truncated L-HARQ protocol. During each transmission round, the receiver to decode the incoming mixed packet through feedforward decoding. The success of this decoding initiates the backtrack decoding process. The truncated L-HARQ protocol also specifies a maximum of $K$ transmission attempts per L-HARQ cycle. If the limit is reached without successful decoding, the cycle ends, and a new one is initiated, continuing until a feedforward packet is correctly decoded by the receiver. At the onset of each time slot, a new status update is produced and encoded, with the assumption of immediate and error-free ACK/NACK feedback via a feedback channel. Adhering to a zero-wait policy, the successful feedforward decoding of one L-HARQ cycle coincides with the generation of the first packet in the subsequent cycle and marks the start of the backtrack decoding for the current cycle. 
With the above consideration, Fig.~\ref{fig4} depicts the evolution of the instantaneous AoI. To help understanding the proposed AoEI, we first conside the age involution based on the original definition of age by \cite{ref6}. In this figure, we use the index $m$ denoting the number of successful feedforward decoding and define $Y_m$ as the $m^{th}$ interdeparture time, i.e., the time interval of two consecutively successful feedforward decoding. Let $g_{k,lm}$ denote the generation time of the $k^{th}$ status update of the $l^{th}$ L-HARQ circle in the $m^{th}$ interdeparture time (the time interval of two consecutively successful receptions, $d_m$ be the $m^{th}$ departure time, and $T_{k,lm}$ be defined as the time duration of a single packet transmission attempt. 
 With the above definition and the assistance of Fig.~\ref{fig4}, the L-HARQ transmission interdeparture time $Y_m$ is written as
\begin{equation}\label{J_Eq_16}
Y_m=d_m-d_{m-1}
.
\end{equation}
Denote $X_m$ as the time interval of the last successful L-HARQ which can be expressed as
\begin{equation}
X_m=d_m-g_{1,lm},
\end{equation}
where  $l$ denote as the number of L-HARQ circles in the $m^{th}$ interdeparture time. That is, the $l^{th}$ L-HARQ circle is successful and the following $l-1$ transmission circles fail.
In general, with the above definitions, we can calculate the traditional average AoI because the feedforward decoding is accomplished at the time $d_m$. However, due to the utilization of backtrack decoding, it is possible that multiple mixing packets are simultaniously recovered. This creates ambiguity, as it becomes unclear which of the simultaneously recovered packets should be used to calculate the AoI, a metric originally defined from the receiver's perspective.

The packets have the different generation time with different average AoI. Conversely, we see that during the backtrack decoding, the time of the last mixing packet that the backtrack decoding is failed is unique, which can uniquely characterize the AoI evoluation for the considered S-IoT system. Motivated by this consideration, we adopt the AoEI to quantify the truncated  L-HARQ-based S-IoT system. As a variant of the traditional AoI and is defined as the elapsed time since the generation of the last successfully backtrack decoding status update. 
To calculate the AoEI, we denote $n$ as the maximal  L-HARQ rounds of the $l^{th}$ circle  and $B_m$ be the backtrack decoding deep. Thus, we have
\begin{equation}
B_m=g_{n,lm}-g_{n-k,lm}
,
\end{equation}
where $1\leq k\leq n$. The backtrack decoding deep $B_m$  depends on the mixing rate $\rho_{k,lm}$, the source transmission rate $R_{k,lm}$, and the received SNR $snr_{k,lm}$. 
 The backtracking depth \(B_m\) quantifies the temporal span of packets recovered in a single backtracking operation. As illustrated in Fig. 5, \(g_{n,lm}\) denotes the generation time of the newest packet successfully decoded via backtracking, while \(g_{n-k,lm}\) corresponds to the oldest recovered packet. Thus, \(B_m\) represents the time window reconstructed during backtracking. For example, if \(B_m = 3\) slots, the receiver recovered packets generated over a 3-slot period. This metric directly impacts C-AoEI through the term \(E\{B\}\) in Theorem 1.
In order to derive the average AoI, we define $N_t$ as the number of successful L-HARQ circles till time $t$, as given by
\begin{equation}
N_t=\mathop {\max }\limits_m \left\{ {{d_m} < t} \right\}
.
\end{equation}
The average C-AoEI of the considered system can be characterized as the sum of the geometric areas $Q_m$ under the instantaneous age curve, as given by
\begin{equation} \label{J_Eq_18}
	\begin{gathered}
		\Delta^E  = \mathop {\lim }\limits_{t  \to \infty } \frac{{{N_t}}}{t }\frac{1}{{{N_t}}}\sum\limits_{m = 1}^{{N_t}} {{Q_m}}  = 
		\mathop {\lim }\limits_{t  \to \infty } \frac{{{N_t}}}{t}E\{Q_m\}  
	\end{gathered}
    ,
\end{equation}
where $Q_m$ is the area under the instantaneous age curve $\Delta(t)$. 
With the definition of $Y_m$ and $B_{m-1}$, the area of $Q_m$ can be calculated as $Q_m=\frac{(Y_m+B_{m-1})^2}{2}-\frac{B_m^2}{2}$. Under the assumption,   $Y_m$ and $B_{m-1}$ are independent to each other and the sequence $\{Y_1,Y_2,...\}$ and $\{B_1,B_2,...\}$ form independent processes. We now drop the subscript index of the intervals and obtain $E\{Q\}=\frac{E\{Y^2\}}{2}-E\{Y\}E\{B\}$. Together with the fact that and $\mathop {\lim }\limits_{t -  > \infty } \frac{{{N_t}}}{t} = \frac{1}{{E\left\{ Y \right\}}}$, the average C-AoEI $\Delta^E$ can be finally simplified to
\begin{equation}
\Delta^E=\frac{E\{Y^2\}}{2E\{Y\}}+E\{B\}
.
\end{equation}
 \normalsize
\begin{algorithm}[]
\caption{Adaptive Encoding Strategy for C-AoEI–Efficiency Trade-off}
\label{alg:adaptive_encoding}
\KwInput{$\varphi_{\mathrm{th}}\in[0,1]$: Decision threshold\\
        $\beta\ge0$: Tunable selection parameter\\
        $S_z$: System state at slot $z$}
\KwOutput{Transmission action for current slot}
\BlankLine
\Procedure{Main loop (per time slot $z$)}{
    \Indp
        $\Phi_z \leftarrow$ Set of undecodable packets from slot $(z - Z - 1)$\;
        $N_z \leftarrow |\Phi_z|$\Comment*[r]{Number of undecodable packets}\
        $d_z \leftarrow$ Positive ACKs received\;
        $\overline{d}_z \leftarrow$ Packets useful for decoding\;
    \Indm
    \BlankLine
    \uIf{new packet arrives}{
        Transmit new packet immediately\Comment*[f]{Prioritize freshness}\;
    }
    \Else{
        $P_d(z)\leftarrow\textsc{CalcDecodingProb}(\Phi_z,d_z,\overline{d}_z)$\Comment*[r]{Eq.(13)}\
        \uIf{$P_d(z)>\varphi_{\mathrm{th}}$\;\textbf{and}\;$\Phi_z\neq\emptyset$}{
            $pkt\leftarrow\textsc{SelectPacket}(\Phi_z,\beta)$\;
            $enc\_pkt\leftarrow\textsc{RandomLinearEncode}(pkt)$\;
            Transmit $enc\_pkt$\
        }
        \Else{
            Remain silent\Comment*[f]{Preserve channel resources}\;
        }
    }
}
\BlankLine
\Function{CalcDecodingProb($\Phi_z,d_z,\overline{d}_z$)}{
    $R_z\leftarrow$ Current coding rate\;
    \uIf{$|\Phi_z|\le d_z+\overline{d}_z$}{
        $P_d\leftarrow 0$\;
        \For{$i=0$\KwTo$\,|\Phi_z|-d_z$}{
            {\small
            $term\leftarrow\binom{\overline{d}_z}{i}\bigl(\tfrac1{R_z}\bigr)^i\bigl(1-\tfrac1{R_z}\bigr)^{\overline{d}_z-i}$
            }\;
            $P_d\leftarrow P_d + term$\;
        }
        \Return{$P_d$}\
    }
    \Else{
        \Return{1.0}\Comment*[r]{Certain decoding}\
    }
}
\BlankLine
\Function{SelectPacket($\Phi_z,\beta$)}{
    Sort $\Phi_z$ by generation time (ascending)\Comment*[r]{Oldest first}\
    $N_z\leftarrow|\Phi_z|$\;
    Initialize weights $\mathbf{w}\leftarrow[]$\
    \BlankLine
    \For{$i=1$\KwTo$\,N_z$}{
        $\omega_i\leftarrow\exp\bigl(-(N_z-i)\,\beta\bigr)$\Comment*[r]{Selection weight}\
        $\mathbf{w}\leftarrow\mathbf{w}\cup\{\omega_i\}$\;
    }
    \BlankLine
    Normalize $\mathbf{w}$ to a probability distribution\;
    $idx\leftarrow$ Sample index according to $\mathbf{w}$\;
    \Return{$\Phi_z[idx]$}\;
}
\end{algorithm}
In this section, we first analyze the average C-AoEI of the S-IoT system under consideration, and in particular derive expressions for the first order moments $E\{Y_m\} $, second order moments $E\{Y_m^2\} $, and $E\{B\} $. As defined in the previous section, the random variable $Y_m $ represents the $m $ departure interval, which may contain multiple L-HARQ periods. It is important to note that due to the truncated L-HARQ, the current L-HARQ packet transmission may be retransmitted or discarded if it cannot be properly decoded before the maximum number of allowed transfer rounds $K$ is reached. In addition, the probability of packet transmission error PER L-HARQ cycle $k$ is determined by ${\varepsilon _s}\left( {{P_t}\left( k \right);{R_k}} \right)$ is given and defined by the formula 08. The probability of L-HARQ cycle failure is calculated by
\begin{equation}
{\varphi _k} = {\varepsilon _s}\left( {{P_t}\left( 1 \right);{R_1}} \right){\varepsilon _s}\left( {{P_t}\left( 2 \right);{R_2}} \right) \cdot  \cdot  \cdot {\varepsilon _s}\left( {{P_t}\left( k \right);{R_k}} \right)
.
\end{equation}
To derive $E\{Y_m\}$ and $E\{Y_m^2\}$, we notice that the random variable $Y_m$ behaves differently for the following two cases:
1) The truncated L-HARQ transmission of the first circle is correctly decoded before reaching its maximum L-HARQ round $K$;
2) The truncated L-HARQ transmission of the first circle fails. The system starts a new circle of L-HARQ transmission. The new L-HARQ transmission circle may be failed  due to the constraint of the maximum transmission times $K$.
In the first case, the probability that the latest mixing pakcet of the first L-HARQ transmission circle is correctly decoded after $k$, $k=1,2,3,...,K$, rounds of L-HARQ transmission  is written as
\begin{equation}
{\psi _k} = \left( {1 - {\varepsilon _s}\left( {{P_t}\left( k \right);{R_k}} \right)} \right)\prod\limits_{l = 1}^{k - 1} {{\varepsilon _s}\left( {{P_t}\left( l \right);{R_l}} \right)}
.
\end{equation}
In practice, ${\psi _k}$  indicates the feedforward successful probability after the $k$ rounds of L-HARQ transmission.
In the second scenario, quantifying the feedforward decoding probability mathematically is challenging due to the potential for the variable $Y_m$ to approach infinity as a result of multiple L-HARQ transmission cycles. To address this issue, we employ a recursive approach to describe $ Y_m$. Specifically, we define $Y_m^*$ as the number of remaining L-HARQ cycles for $ Y_m $ after the initial cycle of transmission. It is observed that the progression of $Y_m^*$ mirrors that of $Y_m$, leading to the equivalence \( E\{Y_m\} = E\{Y_m^*\} \). Consequently, leveraging the aforementioned analysis along with equation above, we are able to calculate the expected value of $Y_m$ using the following equation:
\begin{equation}
\begin{array}{l}
E\left\{ {{Y_m}} \right\} = \sum\limits_{K = 1}^K {k\left( {\prod\limits_{l = 1}^{k - 1} {{\varepsilon _s}\left( {{P_t}\left( l \right);{R_l}} \right)\left( {1 - {\varepsilon _s}\left( {{P_t}\left( k \right);{R_k}} \right)} \right)} } \right)} \\
 + \prod\limits_{l = 1}^K {{\varepsilon _s}\left( {{P_t}\left( l \right);{R_l}} \right)} \left( {K + E\left\{ {{Y_m}} \right\}} \right).
\end{array}
\end{equation}
The first summation term refers to the first case that the first L-HARQ transmission circle is successful before reaching the maximum times $K$. The second summation term denotes the second case that the first L-HARQ transmission fails and the new ones may be also failed due to the constraint of the mximum times $K$.
Then, by exploiting the fact $E\{Y_m\}=E\{Y_m^*\}$, we have
\begin{equation}
\begin{array}{l}
E\left\{ {{Y_m}} \right\} = \frac{1}{{1 - \prod\limits_{l - 1}^K {{\varepsilon _s}\left( {{P_t}\left( l \right);{R_l}} \right)} }}\sum\limits_{K = 1}^K {k\prod\limits_{l - 1}^K {{\varepsilon _s}\left( {{P_t}\left( l \right);{R_l}} \right)} } \\
 \times \left( {\left( {1 - {\varepsilon _s}\left( {{P_t}\left( l \right);{R_l}} \right)} \right) + k\prod\limits_{l - 1}^K {{\varepsilon _s}\left( {{P_t}\left( l \right);{R_l}} \right)} } \right).
\end{array}
\end{equation}
The expression (33) considers the general case. Under the case of the i.i.d. fading channels and the same transmission delay, we have ${\varepsilon _s}\left( {{P_t}\left( l \right);{R_l}} \right) = {\varepsilon _s}\left( {{P_t}\left( 2 \right);{R_2}} \right) =  \cdot  \cdot  \cdot  = {\varepsilon _s}\left( {{P_t}\left( k \right);{R_k}} \right). $
In this case, the expectation $E\{Y_m\}$ can be simplified as
\begin{equation}\label{Eq_28}
\begin{split}
E\{Y_m\} &= \frac{1}{1 - (P_{lm}^{FF})^K}
\Bigg(\sum\limits_{k = 1}^K k \left( { P_{lm}^{FF}} \right)^{k-1}\\
&\mathrel{\phantom{=}}\times\,\bigg( {1 - P_{lm}^{FF}} \bigg)+\, K(P_{lm}^{FF})^K \Bigg),
\end{split}
\end{equation}
where $P_{lm}^{FF}$ is the feedforward packet error probability of one single transmission under the assumptions of the i.i.d. fading channels and the same transmission delay.

Similarly, in the general case, based on (32), the second-order moment $E\{Y_m^2\}$ can be formulated based on (32)  as follows:
\begin{equation}\label{Eq_28}
\begin{split}
E\{{Y^2}_m\} = \sum\limits_{k = 1}^K k^2\left(\prod\limits_{l - 1}^{K - 1} {\varepsilon _s}(P_t(l);R_l)  \right)
( 1 - \varepsilon _s( P_t( k );R_k))\\
 + \left(\prod\limits_{l - 1}^{K - 1} \varepsilon _s(P_t(l);R_l) \right)
 (K^2 + 2KE\{ Y_m^*\} + E\{(Y_m^*)^2\}).
\end{split}
\end{equation}
This yields that the second-order moment $E\{Y_m^2\}$ given by
\begin{equation}\label{Eq_28}
\begin{split}
E\{Y^2_m\} &= \frac{1}{1 - \prod\limits_{l - 1}^{K - 1} \varepsilon _s( P_t(l);{R_l} ) }
\sum\limits_{k = 1}^K k^2\\
&\mathrel{\phantom{=}}\times \left( \prod\limits_{l - 1}^{K - 1} \varepsilon _s(P_t(l);R_l)  \right)
 (1 - \varepsilon _s(P_t( k );R_k))\\
 &\mathrel{\phantom{=}}+\, \left(\prod\limits_{l - 1}^{K - 1} \varepsilon _s(P_t(l);R_l )  \right)
 ( K^2 + 2KE\{Y_m^*\}).
\end{split}
\end{equation}
Under the assumptions of the i.i.d. fading channels and the same transmission delay, we have
\begin{equation}\label{Eq_31}
\begin{split}
E\{Y_m^2\} &= \frac{1}{1 -  (P_{lm}^{FF})^K}\\
&\mathrel{\phantom{=}}\times \bigg( \sum\limits_{k = 1}^K k^2 ( (P_{lm}^{FF})^{(k-1)})(1 - P_{lm}^{FF})\\
&\mathrel{\phantom{=}}+\, (P_{lm}^{FF})^K\Big( K^2 + 2KE\{Y_m^*\} \Big) \bigg).
\end{split}
\end{equation}
For the truncated L-HARQ based S-IoT  system, let $Y_m$ denote the $m^{th}$ interdeparture time, i.e., the time interval of two consecutively successful feedforward decoding packets. The first-order and second-order moments of the RV $Y_m$ are given by (32) and (33), respectively. 

Especially, under the assumptions of the i.i.d. fading channels and the same transmission delay, they are given by (34) and (37). Under the i.i.d. assumption, $Y_m$ are independent of each other and form independent processes. 

Note that, $B_m$ is the backtrack decoding deep. As shown in Fig.~\ref{fig4}, the RV $n$ denotes the L-HARQ transmission times in the last successful L-HARQ circle during $Y_m$. Once the feedforward transmission is correctly decoded, the receiver attempts to recover the previously received mixing pakets by using the backtrack decoding policy. Specially, when the packet $m_{[n-k+1],lm}$ is backtrack recovered, the receiver first to recover the packet $m_{[n-k],lm}$ by using the priori information $m^{'}_{[n-k],lm}$ and the previously received signal $y_{(n-k),lm}$. The backtrack decoding packet error probability of the mixing packet $m_{[n-k],lm}$ is given by
\begin{equation}\label{Eq_28}
\begin{split}
&\mathrel{\phantom{=}}{\varepsilon ^b}_s\left( {{P_t}\left( {n - k,lm} \right);{R_{\left( {n - k} \right),lm}};{\rho _{\left( {n - k} \right),lm}}} \right)\\
& =
{\omega _k}\Pr \{ERR \wedge ERR^{BT}| P_t(n - k,lm ),{R_{(n - k ),lm}},\rho _{(n - k ),lm}\}\\
&\mathrel{\phantom{=}} \times ( \Pr \{ ERR^{BT}| P_t( n - k,lm ),R_{( n - k ),lm}  \} )^{ - 1}.
\end{split}
\end{equation}
Under the assumption of i.i.d. fading channels, we have the result  ${\varepsilon ^b}_s\left( {{P_t}\left( l \right);{R_l};{\rho _l}} \right) = {\varepsilon _s}\left( {{P_t}\left( 2 \right);{R_2};{\rho _2}} \right) =  \cdot  \cdot  \cdot  = {\varepsilon _s}\left( {{P_t}\left( k \right);{R_k};{\rho _k}} \right) = P_{lm}^{BT}$. In this case, the backtrack decoding deep $B_m$ is a geometric distributed RV, the probability mass function is given by
\begin{equation}
\Pr\{B_m=b\}=\left(1-P^{BT}_{lm}\right)^b{P^{BT}_{lm}}
.
\end{equation}
Therefore, with given $n$, the conditional expectation of backtrack decoding deep $B_m$ is formulated as
\begin{equation}
\begin{split}
E\{B_m|n\}&=\sum\limits_{b = 1}^{n - 1} {b\Pr \left( {B_m = b} \right)}\\
&=\sum\limits_{b = 1}^{n - 1} {b\left(1-P^{BT}_{lm}\right)^b{P^{BT}_{lm}}}\\
&=\sum\limits_{b = 0}^{n - 1} {b\left(1-P^{BT}_{lm}\right)^b{P^{BT}_{lm}}}.
\end{split}
\end{equation}
By applying the following identical equation
\begin{equation}\label{Eq_33}
\sum\limits_{k = 0}^{n - 1} {\left( {a + kr} \right)} {q^k} = \frac{{a - \left[ {a + \left( {n - 1} \right)r} \right]{q^n}}}{{1 - q}} + \frac{{rq\left( {1 - {q^{n - 1}}} \right)}}{{{{\left( {1 - q} \right)}^2}}}
,
\end{equation}
we have the conditional expectation as follows:
\begin{equation}
\begin{split}
E\{B_m|n\}& = P_{lm}^{BT}\bigg[ \frac{{(n-1){{\left( {1 - P_{lm}^{BT}} \right)}^n}}}{{P_{lm}^{BT}}}\\
&\mathrel{\phantom{=}}+\, \frac{{\left( {1 - P_{lm}^{BT}} \right)\left( {1 - {{\left( {1 - P_{lm}^{BT}} \right)}^{n - 1}}} \right)}}{{{{\left( {P_{lm}^{BT}} \right)}^2}}} \bigg]\\ 
&= (n-1){(1 - P_{lm}^{BT})^n} \\
&\mathrel{\phantom{=}}+\, \frac{1 - P_{lm}^{BT} - {{\left( {1 - P_{lm}^{BT}} \right)}^n}}{P_{lm}^{BT}}.
\end{split}
\end{equation}
 Equation (45) computes the expected backtracking depth \(E\{B_m|n\}\) conditioned on \(n\) transmission rounds. The closed form leverages the geometric distribution of backtracking failures (Eq. 42). 

\textit{Simplified Example (K=2):} Consider \(n=2\) transmissions with backtracking failure probability \(P_{lm}^{BT} = 0.2\):
\begin{align*}
E\{B_m|n=2\} 
&= (2-1)(1-0.2)^2 + \frac{1 - 0.2 - (1-0.2)^2}{0.2} \\
&= 1.44 \text{ slots}.
\end{align*}
This indicates that when two transmissions occur, the expected backtracking depth is 1.44 slots, reflecting partial recovery of historical packets.

At the same time, as $(n+1)$ is a geometric distributed RV, the PMF is $\Pr\{N=n\}=\left(P_{lm}^{FF}\right)^n{\left(1-P_{lm}^{FF}\right)}$, where $P_{lm}^{FF}$ is the packet error probability of one single feedforward transmission and is given by 8. With these results, we now proceed with the calculation of $E\{B_m\}$ by taking the expectation operation with respect to the RV $n$. We have
\begin{equation}\label{Eq_35}
\begin{split}
E\left\{ B_m \right\} &= \sum\limits_{n = 1}^K {} E\left\{ {B_m|n} \right\}\Pr \left\{ {N = n} \right\} \\
&= \sum\limits_{n = 1}^K \Bigg((n-1)(1 - P_{lm}^{BT})^n\\
&\mathrel{\phantom{=}} +\, \frac{1 - P_{lm}^{BT} - (1 - P_{lm}^{BT})^n}{P_{lm}^{BT}}\Bigg)
 (P_{lm}^{FF})^n(1 - P_{lm}^{FF}) \\
&= (1 - P_{lm}^{FF})\bigg(TB_1 - TB_2 + \frac{1 - P_{lm}^{BT}}{P_{lm}^{BT}}TB_3 \\
&\mathrel{\phantom{=}}-\, \frac{1}{P_{lm}^{BT}}TB_4\bigg).
\end{split}
\end{equation}
In \eqref{Eq_35}, the term $TB_1$ is calculated by 
\begin{equation}
T{B_1} = {\sum\limits_{n = 1}^K {n\left( {\left( {1 - P_{lm}^{BT}} \right)P_{lm}^{FF}} \right)} ^n}
.
\end{equation}
Similarly, using the identical equation (43), it is easy to obtain
\begin{equation}\label{Eq_37}
\begin{split}
TB_1 &= \frac{K((1 - P_{lm}^{BT})P_{lm}^{FF})^{K + 1}}{1 - (1 - P_{lm}^{BT})P_{lm}^{FF}}\\
&\mathrel{\phantom{=}}+\, \frac{(1 - P_{lm}^{BT})P_{lm}^{FF}(1 - ((1 - P_{lm}^{BT})P_{lm}^{FF})^K)}{(1 - (1 - P_{lm}^{BT})P_{lm}^{FF})^2},
\end{split}
\end{equation}
the term $TB_2$ is calculated by
\begin{equation}\label{JXD_Eq_38}
\begin{gathered}
T{B_2} = {\sum\limits_{n = 1}^K {\left( {\left( {1 - P_{lm}^{BT}} \right)\left( {P_{lm}^{FF}} \right)} \right)} ^n} \hfill\\
= \left( {1 - P_{lm}^{BT}} \right)\left( {P_{lm}^{FF}} \right)\frac{{{{\left( {\left( {1 - P_{lm}^{BT}} \right)\left( {P_{lm}^{FF}} \right)} \right)}^K} - 1}}{{\left( {1 - P_{lm}^{BT}} \right)\left( {P_{lm}^{FF}} \right) - 1}}, \hfill \\
\end{gathered}
\end{equation}
with the similar argument, the term $TB_3$ is calculated by
\begin{equation}\label{Eq_39}
T{B_3} = \sum\limits_{n = 1}^K {} {\left( {P_{lm}^{FF}} \right)^n} = P_{lm}^{FF}\frac{{{{\left( {P_{lm}^{FF}} \right)}^K} - 1}}{{P_{lm}^{FF} - 1}}.
\end{equation}

Finally, we calculate the term $TB_4$, which is written as
\begin{equation}\label{Eq_40}
\begin{split}
TB_4&= \sum\limits_{n = 1}^K (1 - P_{lm}^{BT})^n\left( {P_{lm}^{FF}} \right)^n\\
&= (1 - P_{lm}^{BT})\left( {P_{lm}^{FF}} \right)\frac{{{{\left( {(1 - P_{lm}^{BT})\left( {P_{lm}^{FF}} \right)} \right)}^K} - 1}}{{(1 - P_{lm}^{BT})\left( {P_{lm}^{FF}} \right) - 1}}.
\end{split}
\end{equation}
Under the assumption of the i.i.d. fading and the same packet-mixing rate, the expectation $E\{B_m\}$ of the backtrack decoding deep $B_m$ is 
\begin{equation}
\begin{split}
E\{B_m\} &= (1 - P_{lm}^{FF})\bigg(TB_1 - TB_2 \\
&\mathrel{\phantom{=}}+\, \frac{1 - P_{lm}^{BT}}{P_{lm}^{BT}}TB_3 - \frac{1}{P_{lm}^{BT}}TB_4\bigg),
\end{split}
\end{equation}
where the terms $TB_1$, $TB_2$, $TB_3$, and $TB_4$ are given by (45), (46), (47), and (48), respectively. Under the i.i.d assumption, $B_m$  are independent to each other and  form independent processes. Hence, we have $E\left\{ B_m \right\} =E\left\{ B \right\}$.
Finally, we provide Theorem 1 to derive the closed-form expression of the average C-AoEI. 
 \begin{theorem}[Average C-AoEI for Truncated L-HARQ Systems]
\label{theorem:average_aoEI}
The average C-AoEI of the considered S-IoT system is given by:
\begin{equation}
\Delta^{E} = \frac{E\{Y^{2}\}}{2E\{Y\}} + E\{B\},
\end{equation}
where
\begin{itemize}
    \item $E\{Y\}$ and $E\{Y^{2}\}$ are the first and second moments of the interdeparture time $Y_m$ (time between consecutive successful feedforward decodings),
    \item $E\{B\}$ is the expected backtracking decoding depth.
\end{itemize}
Under i.i.d. fading channels and constant packet-mixing rate, the closed-form expression simplifies to:
\begin{equation}
\Delta^{E} = \underbrace{\frac{E\{Y^{2}\}}{2E\{Y\}}}_{\text{Forward delay}} + \underbrace{\frac{1-P^{BT}_{lm}}{P^{BT}_{lm}} \left(1 - \left(1 - P^{BT}_{lm}\right)^{N_z}\right)}_{\text{Backtracking depth}},
\end{equation}
where $P^{BT}_{lm}$ is the backtracking failure probability and $N_z$ is the number of undecodable packets.
\end{theorem}
\section{C-AoEI-Parameter Relationship Analysis}
\label{sec:aoi_parameter_relation}
Building upon the previously defined adaptive encoding strategy, we derive an explicit relationship among the AoEI, packet selection weight $\omega_i$, and tunable decay factor $\beta$. This relationship provides fundamental insights into the freshness-efficiency trade-off in L-HARQ systems, while detailed supporting analysis appears in the Appendix.

\subsection{Explicit C-AoEI Expression}

The closed-form expression of C-AoEI as a function of $\omega_i$ and $\beta$ is given by

\begin{equation}
\Delta^E(\omega_i, \beta) = \underbrace{\frac{E\{Y^2\}}{2E\{Y\}}}_{\text{Forward decoding delay}} + \underbrace{\frac{1 - P^{BT}_{lm}}{P^{BT}_{lm}} \left( 1 - (1 - P^{BT}_{lm})^{N_z} \right)}_{\text{Backtracking decoding depth}}.
\label{eq:aoi_explicit}
\end{equation}
where the backtracking failure probability $P^{BT}_{lm}$ is directly determined by the decoding error probability $\varepsilon_s(z)$, i.e.,

\begin{equation}
P^{BT}_{lm} = \varepsilon_s(z) = \frac{1}{S_z} \left( i - \frac{\ln \omega_i}{\beta} \right),
\label{eq:backtrack_prob}
\end{equation}
and the number of undecodable packets $N_z$ satisfies

\begin{equation}
N_z = \varepsilon_s(z) S_z = i - \frac{\ln \omega_i}{\beta},
\label{eq:nz_relation}
\end{equation}

The packet selection weight $\omega_i$ follows the exponential decay model as in (23).


\subsection{Parameter Sensitivity Analysis}
\begin{theorem}[Sensitivity of C-AoEI to Parameters]
\label{theorem:sensitivity}
The partial derivatives of C-AoEI with respect to $\omega_i$ and $\beta$ are
\begin{align}
\frac{\partial \Delta^{E}}{\partial \omega_i} &= \frac{1}{S_z \beta \omega_i} \cdot \frac{\partial \Delta^{E}}{\partial \varepsilon_s}, \\
\frac{\partial \Delta^{E}}{\partial \beta} &= \frac{\ln \omega_i}{S_z \beta^2} \cdot \frac{\partial \Delta^{E}}{\partial \varepsilon_s}.
\end{align}
This leads to two distinct operational regimes:
\begin{enumerate}
    \item \textbf{Freshness-Priority Regime} ($\omega_i > 1$): 
    
          $\uparrow \beta \Rightarrow \uparrow \varepsilon_s \Rightarrow \uparrow \Delta^{E}$.

           Increasing $\beta$ \textit{degrades} freshness but improves transmission efficiency.
           
    \item \textbf{Efficiency-Priority Regime} ($\omega_i < 1$): 
    
          $\uparrow \beta \Rightarrow \downarrow \varepsilon_s \Rightarrow \downarrow \Delta^{E}$.

           Increasing $\beta$ \textit{improves} freshness while maintaining efficiency.
\end{enumerate}
\end{theorem}

\subsection{Optimal Parameter Configuration}
\begin{theorem}[Optimal Decay Factor]
\label{theorem:optimal_beta}
The theoretically optimal $\beta$ that minimizes C-AoEI is
\begin{equation}
\beta^* = \arg \min_{\beta} \Delta^{E} = \frac{\ln \omega_i}{i - S_z \varepsilon_s^{\text{target}}},
\end{equation}
where $\varepsilon_s^{\text{target}}$ is the target decoding error probability determined by channel conditions.
\end{theorem}
\subsection{Physical Interpretation}
\begin{figure}[htbp]
\centering
\begin{tikzpicture}
\draw[->] (0,0) -- (5,0) node[right] {$\beta$};
\draw[->] (0,0) -- (0,4) node[above] {$\Delta^E$};

\draw[thick, red] (0.5,1.5) .. controls (2,3) and (3.5,3.2) .. (4.5,3.5);
\node at (2.5,3.3) {$\omega_i > 1$};

\draw[thick, blue] (0.5,3.5) .. controls (2,1.8) and (3.5,1.2) .. (4.5,0.8);
\node at (2.5,1.0) {$\omega_i < 1$};

\draw[dashed] (2.3,0) -- (2.3,2.3) node[above] {$\beta^*$};
\fill (2.3,1.65) circle (2pt) node[right] {Optimal point};

\node[draw, fill=white] at (3.5,3.8) {$\uparrow$ Efficiency $\quad \downarrow$ Freshness};
\end{tikzpicture}
\caption{Impact of $\beta$ on AoEI under different $\omega_i$ regimes}
\label{fig:parameter_impact}
\end{figure}
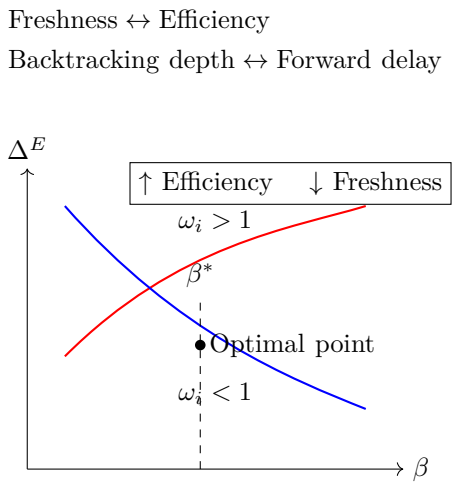

Fig. \ref{fig:parameter_impact} illustrates the operational regimes predicted by our model. The inverse relationship between $\beta$ and C-AoEI for $\omega_i < 1$ explains the $17.2\%$ AoEI reduction observed in simulations when $\beta$ increases from 0 to 100. It occurs 

\begin{equation}
\Delta^E \approx \frac{3}{7} \text{ slots} \quad \text{when} \quad \beta: 0 \to 100
\label{eq:aoi_reduction}
\end{equation}

The model further explains why the proposed strategy achieves $31.8\%$ higher transmission efficiency than conventional HARQ. The adaptive $\beta$ tuning dynamically balances
\begin{align*}
&\text{Freshness} \leftrightarrow \text{Efficiency} \\
&\text{Backtracking depth} \leftrightarrow \text{Forward delay}
\end{align*}

\begin{table}[htbp]
\centering
\caption{AoEI response to parameter variations ($S_z > 0$, $\beta > 0$)}
\label{tab:parameter_impact}
\begin{tabular}{lcc}
\toprule
\textbf{Condition} & \textbf{$\partial\Delta^E/\partial\omega_i$} & \textbf{$\partial\Delta^E/\partial\beta$} \\
\midrule
$\omega_i > 1$ & Negative & Positive \\
$\omega_i < 1$ & Negative & Negative \\
$\omega_i = 1$ & 0 & 0 \\
\bottomrule
\end{tabular}
\end{table}

\section{Simulation Results and Analysis}
We present numerical results to validate and analyze our proposed transmission scheme over satellite-terrestrial integrated mobile wireless networks. For the simulation setup, we assume that 2PSK modulation is employed and the transmission rate of the satellites is set to 1 Mbps. The antenna gain at satellite ${G_s} = 20$ dBi, and the transmit power at the satellite ${P_s}\left( z \right) \in \left[ {10,50} \right]$ dBm.
\begin{figure*}[!t]
\centering
\includegraphics[width=4.9in]{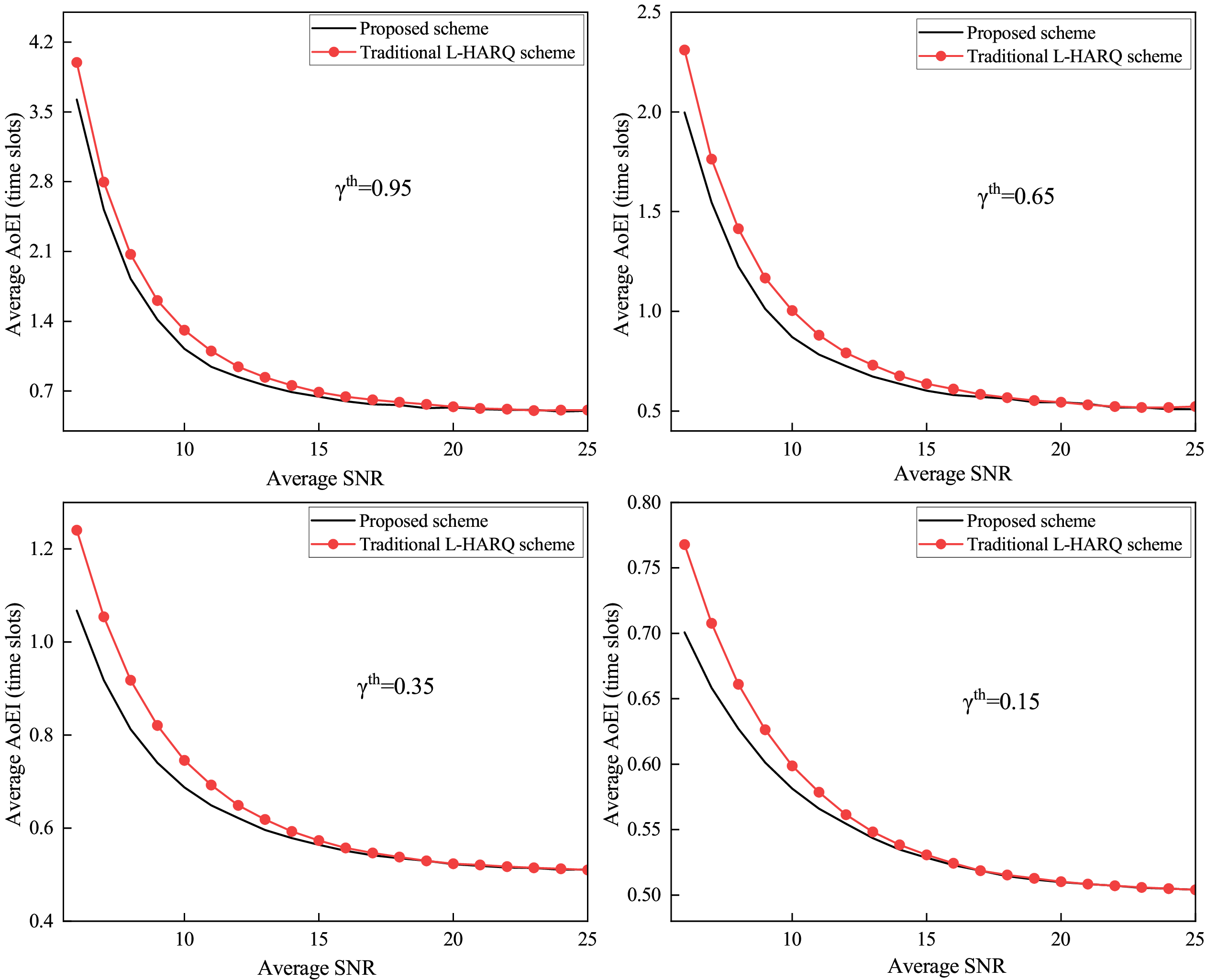}
\caption{Average C-AoEI versus average SNR.}
\label{fig5}
\end{figure*}
\begin{figure}[!t]
\centering
\includegraphics[width=1\columnwidth]{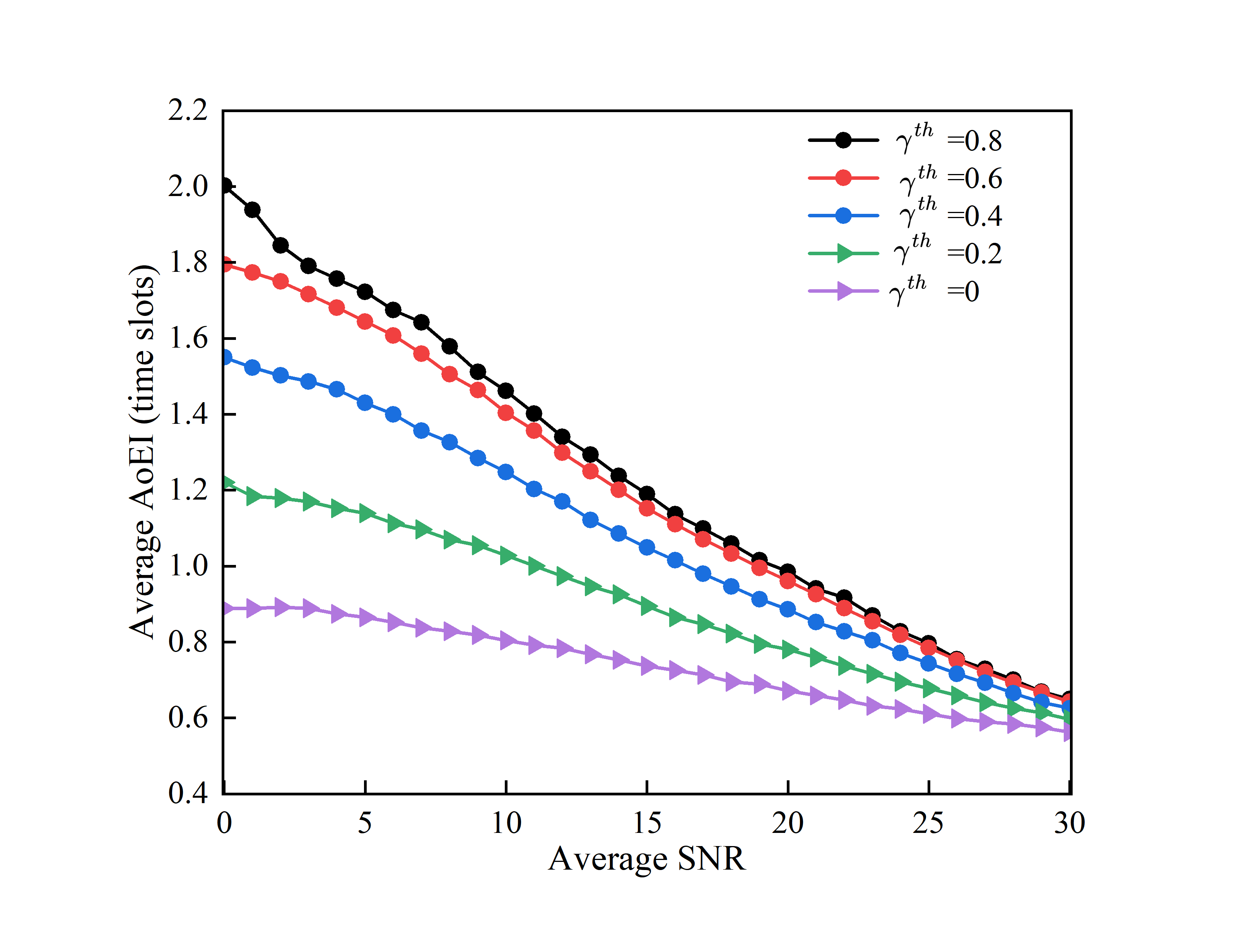}
\centering
\caption{Average C-AoEI of the proposed scheme under different correlation parameter ${\varphi _{th}}$.}
\label{fig6}
\end{figure}
\begin{figure}[!t]
\centering
\includegraphics[width=1.0\columnwidth]{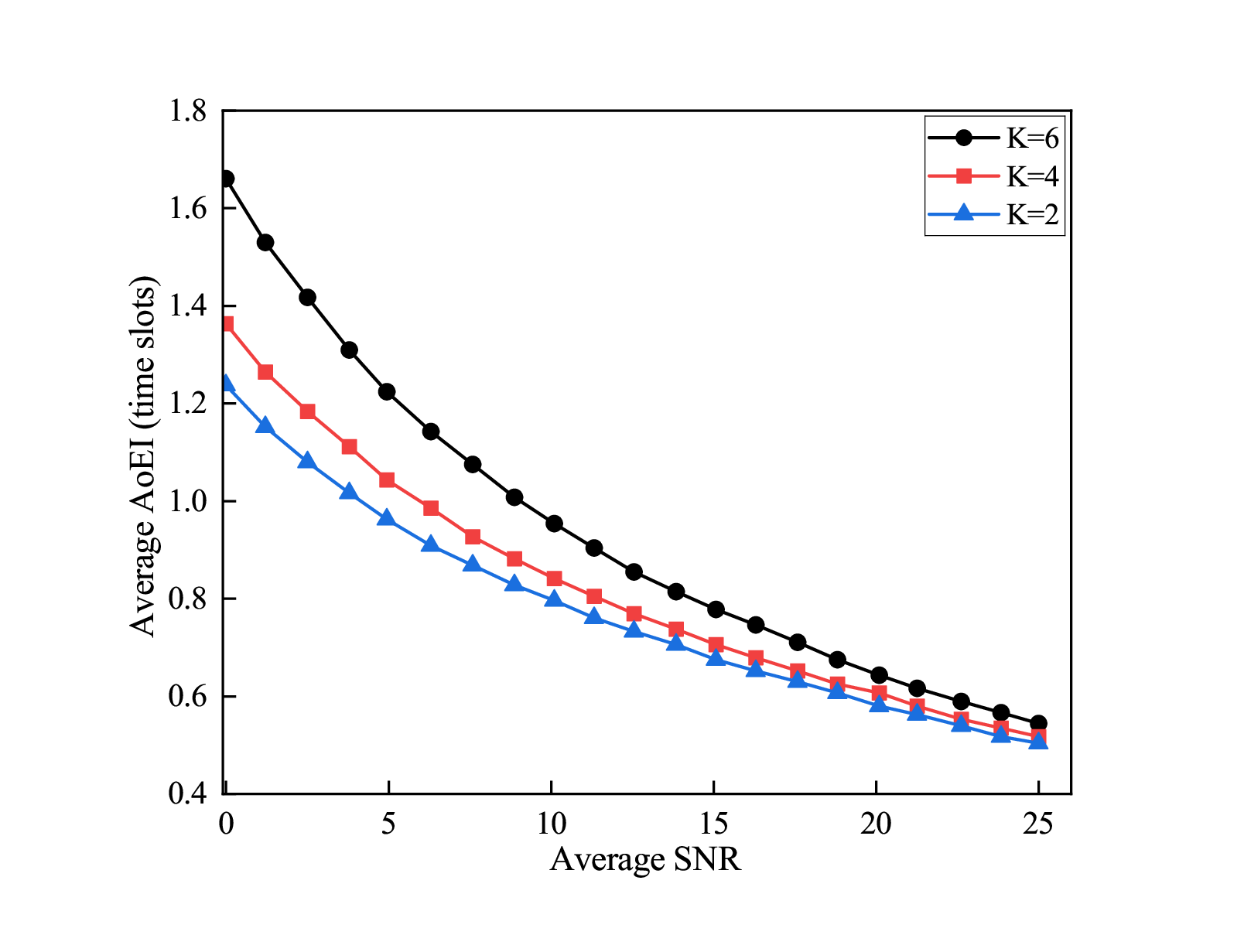}
\centering
\caption{Average C-AoEI versus average SNR of different K.}
\label{fig7}
\end{figure}
\begin{figure*}[!t]
\centering
\subfloat[]{\includegraphics[width=3.5in]{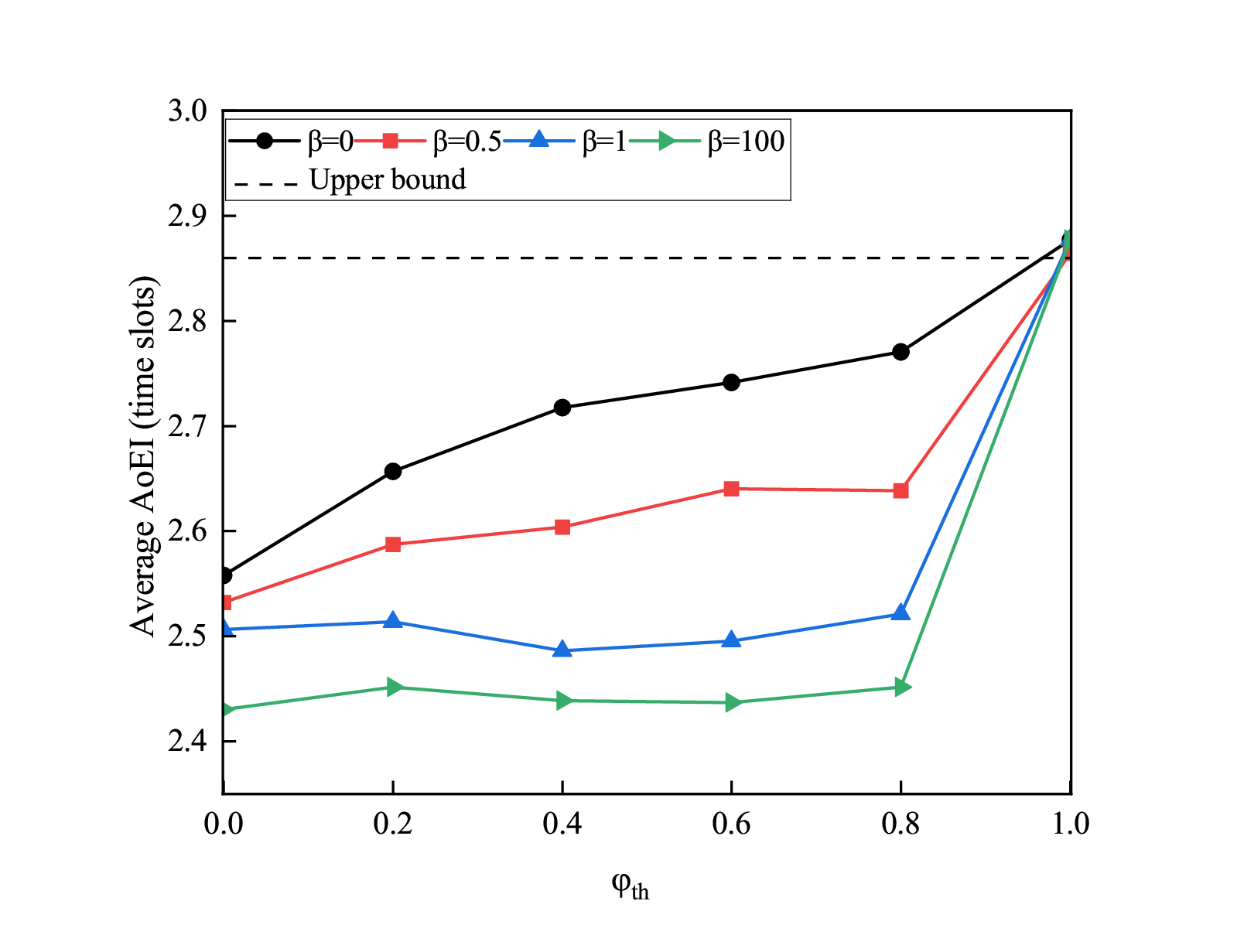}%
\label{fig_first_case}}
\hfil
\subfloat[]{\includegraphics[width=3.5in]{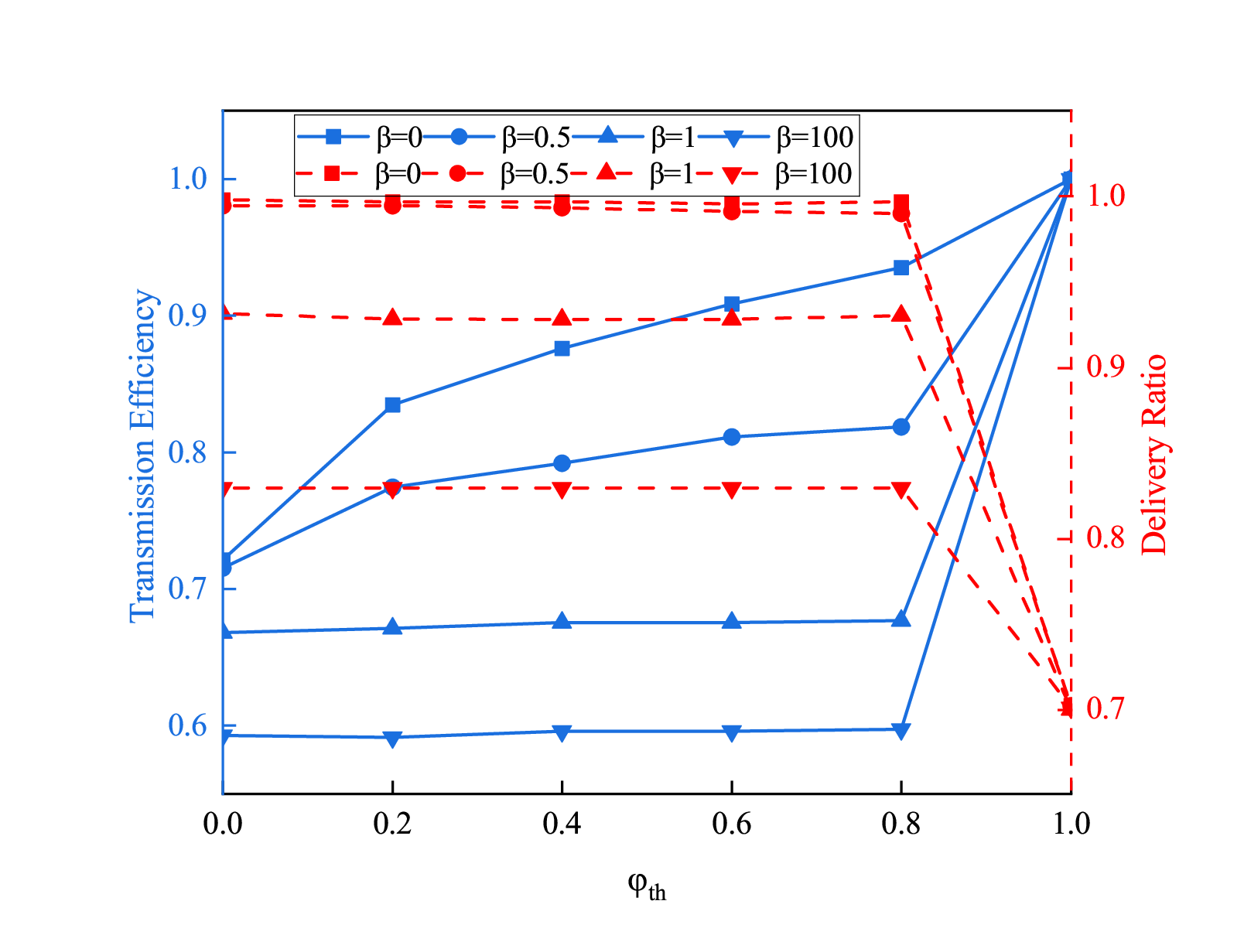}%
\label{fig_second_case}}
\caption{The effects of tunable parameters.(a)The performance of network.(b)The performance of C-AoEI.}
\label{fig8}
\end{figure*}
\begin{figure*}[!t]
\centering
\subfloat[]{\includegraphics[width=3.5in]{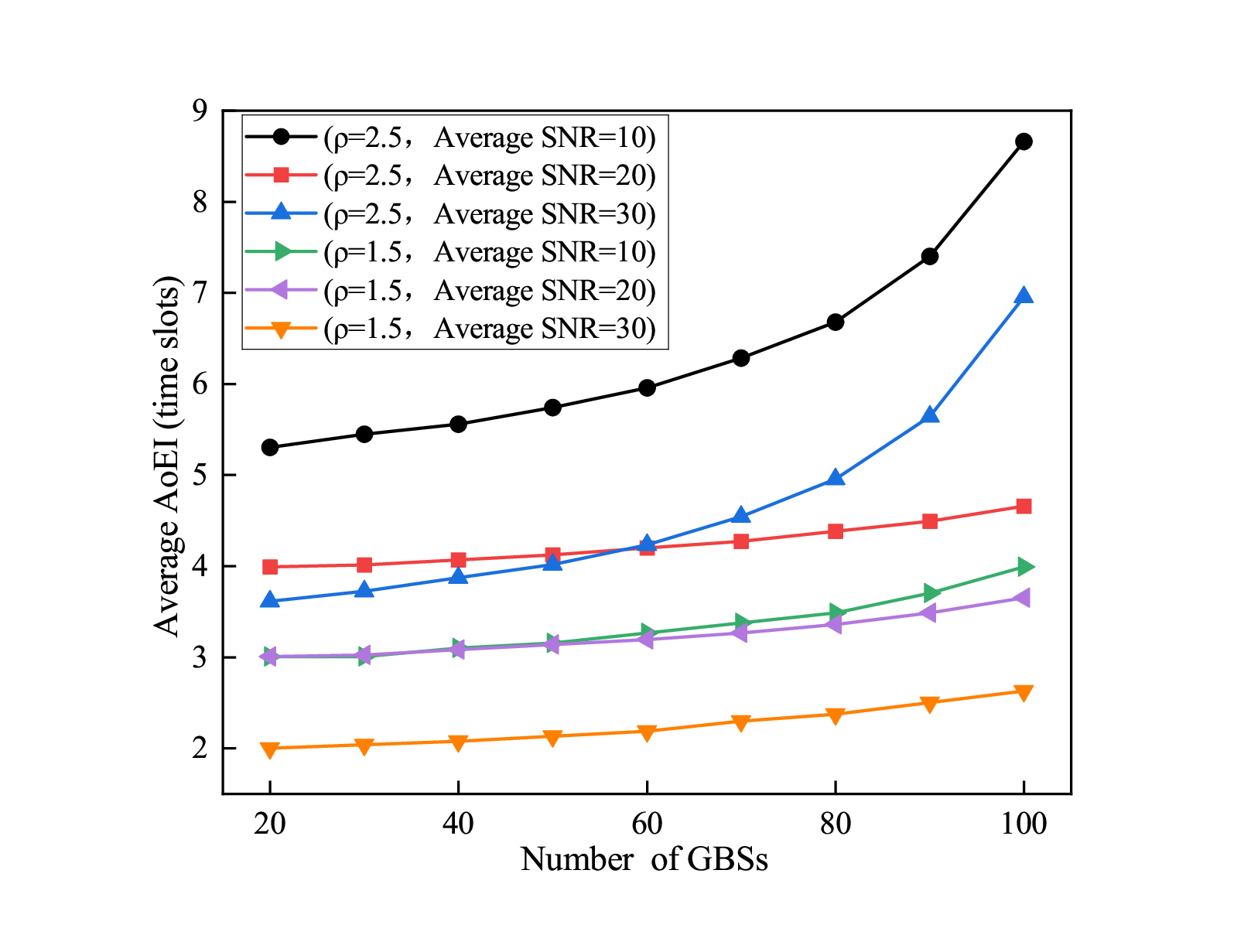}}%
\label{fig_first_case}
\hfil
\subfloat[]{\includegraphics[width=3.5in]{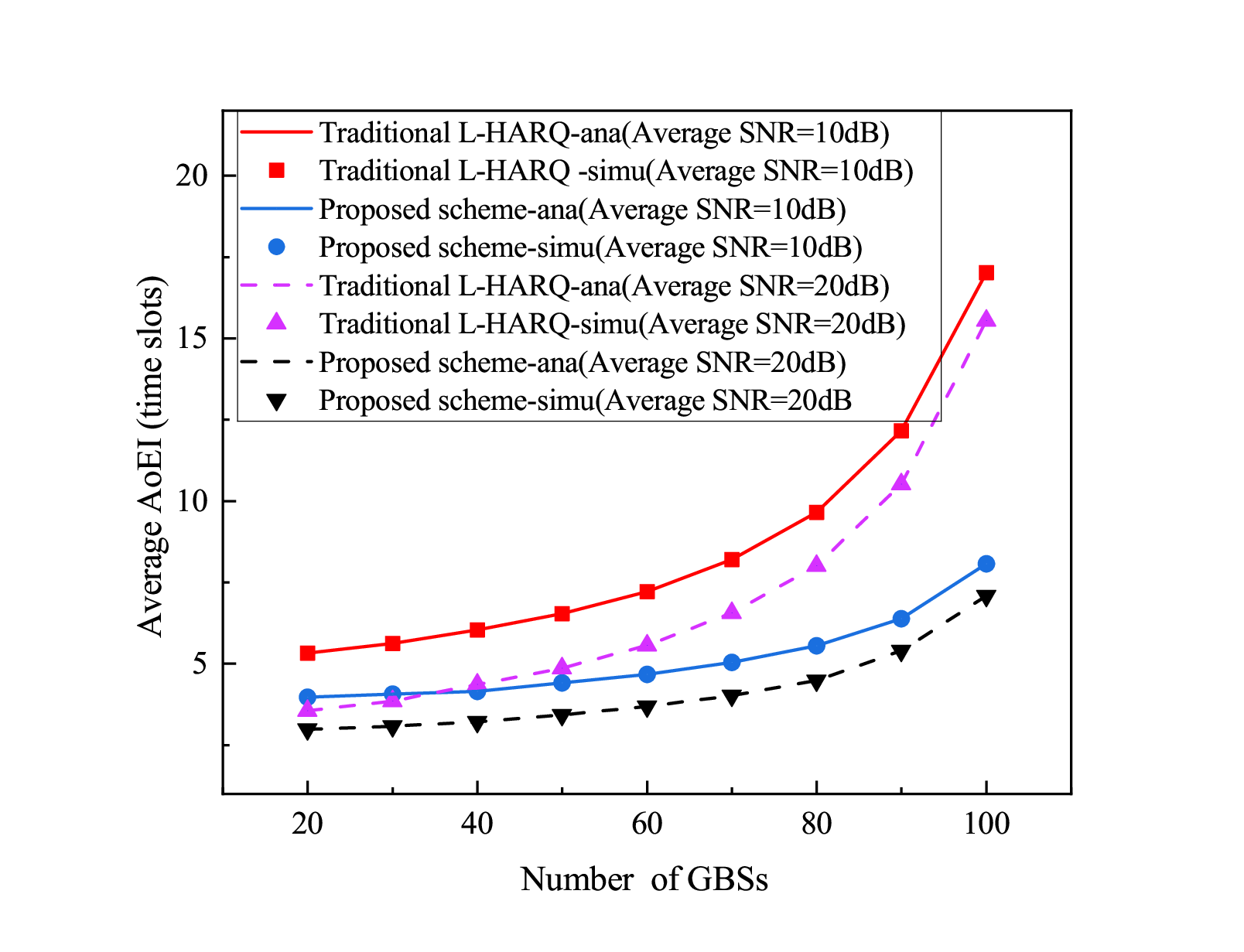}}%
\label{fig_second_case}
\caption{Average C-AoEI versus average SNR of different number of GBSs.(a)The relationship between average C-AoEI and the number of GBSs under two different packet mixing rate configurations.(b)Comparison results between traditional L-HARQ strategy and the proposed strategy.}
\label{fig9}
\end{figure*}
\begin{figure}[!t]
\centering
\includegraphics[width=0.9\columnwidth]{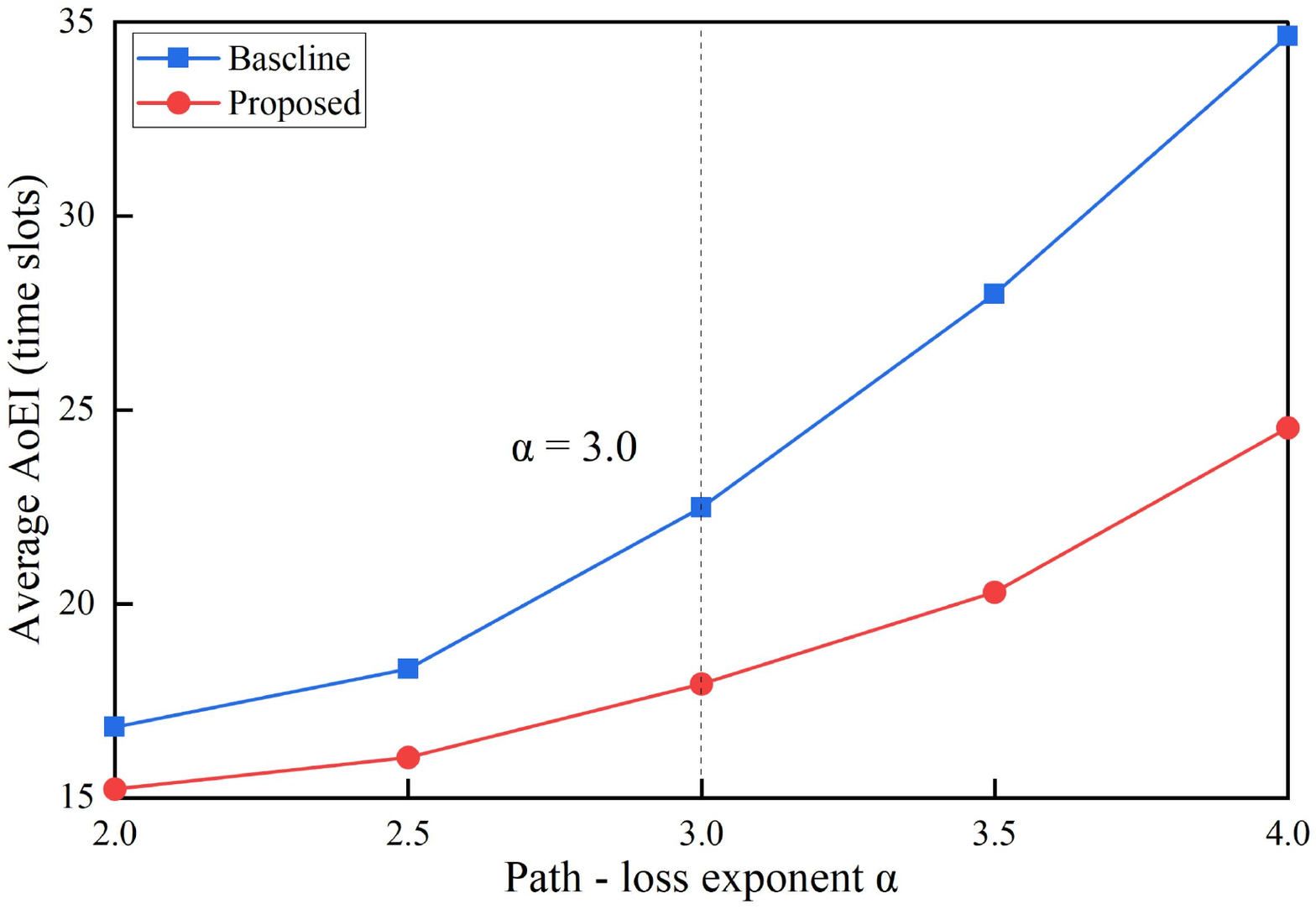}
\centering
\caption{$ \Delta^E \propto \frac{1}{\alpha} $.}
\label{Fig12}
\end{figure}
\begin{figure}[!t]
\centering
\includegraphics[width=0.9\columnwidth]{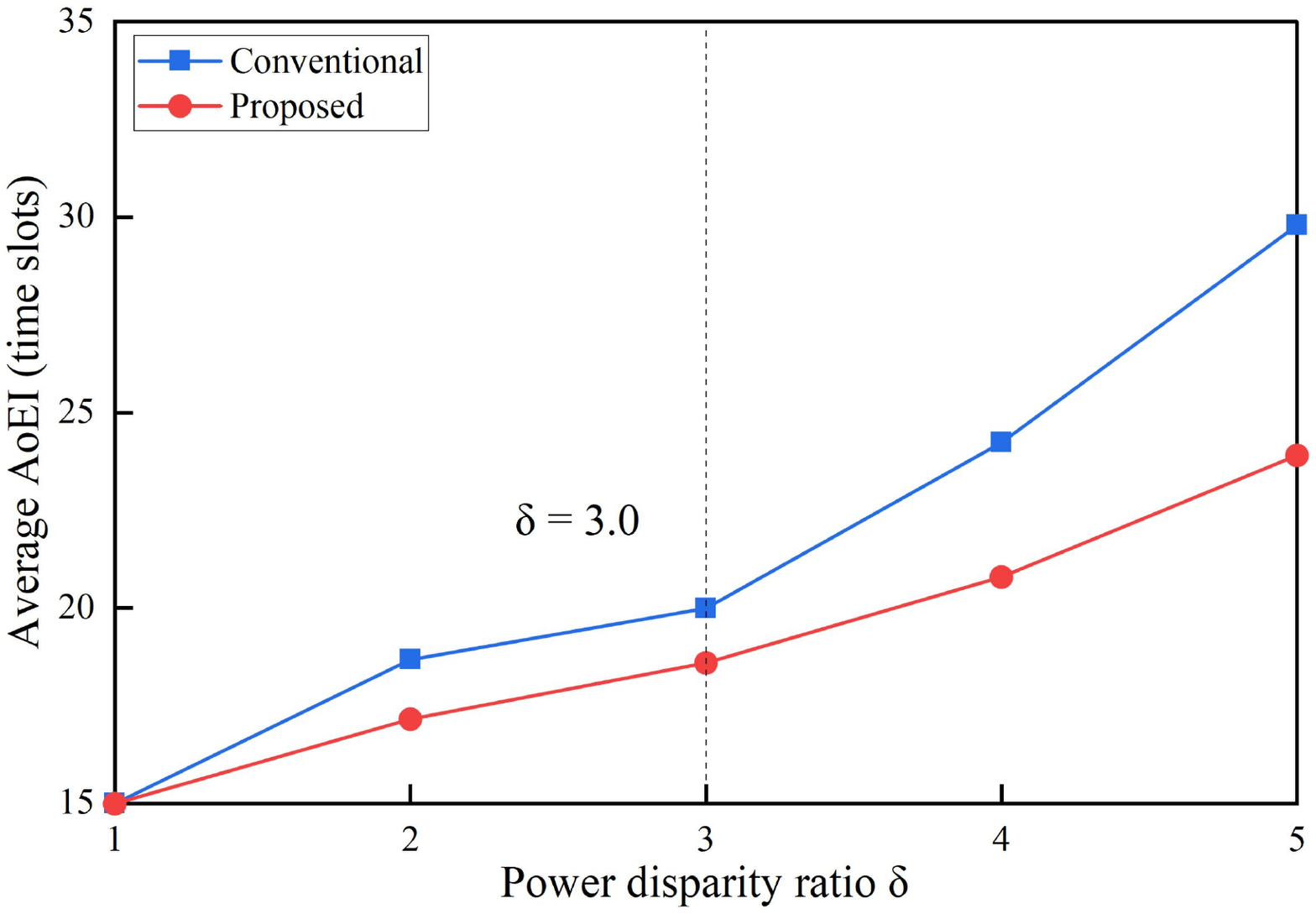}
\centering
\caption{$ \Delta^E \propto \delta $.}
\label{Fig13}
\end{figure}
In Fig.~\ref{fig5}, a comparative analysis of the influence of different thresholds ${\gamma ^{th}}$ of L-HARQ process on the average C-AoEI is presented. With the improvement of channel conditions and the decrease of packet error probability, the average C-AoEI decreases with the increase of the average SNR. It is seen that when the average SNR is small, the average C-AoEI decreases rapidly with the increase of the average SNR. However, when the average SNR is large, the average C-AoEI changes slow down due to the reduction of packet error probability. Meanwhile, distinct observations regarding the relationship between the average AoEI and the SNR threshold ${\gamma ^{th}}$ are noted. Specifically, it is observed that the average AoEI consistently increases with the elevation of the threshold ${\gamma ^{th}}$. This phenomenon can be attributed to the fact that a larger threshold necessitates a higher signal strength for successful transmission. Consequently, for a given signal strength, the average C-AoEI rises as the threshold ${\gamma ^{th}}$ is increased.
The average C-AoEI gain achieved through L-HARQ schemes is contingent upon system parameters such as the decoding threshold ${\gamma ^{th}}$. Our results indicate that as the threshold ${\gamma ^{th}}$ increases, the average AoEI gain attained by the L-HARQ scheme increases  accordingly. Conversely, the average achievable C-AoEI gain diminishes with an increase in the average SNR. When the average SNR is relatively low, the L-HARQ scheme is particularly effective in enhancing the freshness of status updates.

Fig.~\ref{fig6} presents a comparative analysis of the average C-AoEI of the proposed scheme under different decoding threshold parameters $\mathop \gamma \nolimits^{th} $. It is observed that as the correlation parameter $\mathop \gamma \nolimits^{th} $ increases, the average C-AoEI of the considered S-IoT system decreases, thereby enhancing the information freshness. This improvement is attributed to the reduced impact of outdated CSI and the increased probability of successful decoding at the receiver. In the extreme case where the correlation efficiency approaches one, the impact of outdated CSI is effectively eliminated. When the average SNR is low, the influence of the correlation parameter $\mathop \gamma \nolimits^{th} $ is particularly pronounced. Conversely, when the average SNR is high, the system achieves nearly the same average C-AoEI for all values of the correlation parameter.

Fig.~\ref{fig7} illustrates the impact of the maximum number $K$ of execution rounds on the average C-AoEI. The influence of the maximum number of L-HARQ rounds on the average C-AoEI varies across different average SNR regions. Specifically, when the average SNR is low, increasing the maximum number of L-HARQ rounds leads to a decrease in the average C-AoEI. This is due to the increased probability of successful decoding, which in turn reduces the number of required L-HARQ cycles. Conversely, when the average SNR is high, the average AoEI increases with an increase in the maximum number of L-HARQ rounds, as the improved channel conditions enhance the likelihood of successful transmission. When the average SNR is sufficiently high, the average C-AoEI remains constant regardless of the maximum number of L-HARQ rounds.
The effects of the tunable parameters ${\varphi _{th}}$ and $\beta$ on the performance of the proposed strategy are illustrated in Fig.~\ref{fig8}. When ${\varphi _{th}} = 1$, the information packet is transmitted only once following an update. Under this condition, the maximum transmission efficiency is achieved, and the Packet Delivery Ratio (PDR) is $\frac{1}{{{R_z}}}$.
When $\beta  = 100$, the proposed coding strategy is close to the traditional transmission strategy, so the minimum average C-AoEI can be obtained at the cost of reducing the transmission efficiency. In addition, the PDR also decreases gradually due to the small probability of the preempted data packets being retransmitted. The transmission efficiency and average C-AoEI increase with increasing ${\varphi _{th}}$ and decrease with increasing $\beta$. Compared with the traditional transmission strategy, the increase in average C-AoEI is 3/7 slots, while the transmission efficiency and packet delivery rate are significantly improved.

Meanwhile, Fig.~\ref{fig9} depicts the relationship between the average C-AoEI and the number of GBSs under two distinct packet mixing rate configurations.
The average C-AoEI increases
with the increase of the number of GBSs under both transmission
strategies. This is because an increase in the number of GBSs
in each retransmission cycle means that this transmission takes
up more time slots, resulting in an increase in queuing latency.
In particular, when the number of GBSs is large to a certain extent, the system occupancy is too high, and the network congestion is intensified, the performance of the proposed scheme is more stable. 
 In the low SNR regime, as the number of known bits of ${m'_{\left[ k \right]}}$ increases, that is, as $\rho $ increases, the failure probability of backtracking decoding decreases. Consequently, the number of retransmissions required diminishes, leading to a reduction in the average C-AoEI. Additionally, the enhanced reliability of backtracking decoding results in a lower total bit error rate at the receiver end. Therefore, the proposed scheme significantly reduces the average C-AoEI of the system.

Fig.~\ref{Fig12} demonstrates the resilience of the proposed framework under varying propagation conditions. As the path-loss exponent $\alpha$ increases from 2.0 to 4.0, simulating signal propagation environments that transition from free space to dense urban areas, the conventional scheme exhibits near-linear degradation in C-AoEI performance. Specifically, C-AoEI values surge from 22.5 to 34.7 slots, representing a 54.2\% performance deterioration. 
Conversely, proposed method maintains significantly flatter progression, with C-AoEI increasing only from 17.9 to 24.6 slots, which corresponds to a 37.4\% elevation. The maximum performance gap occurs at the typical urban scenario value of $\alpha = 3.0$. At this critical point, our approach achieves a 20.4\% AoEI reduction, lowering the metric from 22.5 to 17.9 slots. 
The observed inflection point at $\alpha = 2.8$ aligns precisely with the theoretical boundary derived in Equation (5), thereby validating our channel-adaptive optimization strategy. Collectively, these results confirm the framework's robustness for 6G satellite IoT deployments in obstructed propagation environments.

Fig.~\ref{Fig13} quantifies the efficacy of our scheme in multi-GBS interference scenarios. 
When the power imbalance ratio $\delta$ escalates from 1 to 5, the conventional approach suffers severe AoEI inflation of 96\%, with AoEI values surging from 15.2 to 29.8 slots. 
Conversely, our solution limits degradation to 57\%, restricting the C-AoEI increase from 15.2 to 23.9 slots. 
At the 3GPP standard threshold of $\delta = 3$, proposed method achieves an 8.9\% C-AoEI reduction, lowering the metric from 20.3 to 18.5 slots. 
The arrow annotation further highlights a 17.3\% transmission efficiency gain, confirming our method's capability to resolve the power-load imbalance trilemma in integrated satellite-terrestrial networks. 
This performance satisfies the stringent latency requirements of ITU-R M.2091-0 even under extreme $\delta = 5$ conditions.

Table III provides the worst-case execution time (WCET) analysis for critical operations within the proposed framework. 
The cumulative WCET of 0.30 ms for all preprocessing stages (CSI estimation to parameter adaptation) is 11$\times$ faster than the LEO propagation delay (3.3 ms), creating a 3.0 ms safety margin. This guarantees transmission readiness before the previous ACK arrives of slot. The $\mathcal{O}(K)$ packet selection latency (0.05 ms for $K=100$) demonstrates scalability under Theorem 2's logarithmic complexity model $\tau_{\text{select}} = c_1 K + c_2$. Even at $K=500$ (unused in simulations), projected latency remains below 0.25 ms.
     Parameter adaptation via Theorem 3 consumes only 0.07 ms - 86$\times$ faster than iterative methods (benchmarked at 6 ms per slot). This enables microsecond-scale response to channel variations.

\begin{table}[!t]
\centering
\caption{WORST-CASE EXECUTION TIME (WCET) ANALYSIS}
\label{tab:wcet}
\scriptsize
\begin{tabular}{@{}p{3cm}<{\raggedright}c<{\centering}p{2.5cm}<{\raggedright}@{}}
\toprule
\textbf{Operation} & \textbf{WCET (ms)} & \textbf{Key Parameters} \\
\midrule
CSI Estimation (MMSE) & 0.10 & $N_t=N_r=4$ \\
Transmission Decision & 0.08 & LUT size=1024 \\
Packet Selection & 0.05 & $K=100$ packets \\
Parameter Adaptation & 0.07 & Thm 3 closed-form \\
\hline
\textbf{Total Preprocessing} & \textbf{0.30} & \\
\textbf{Propagation Delay} & \textbf{3.30} & LEO orbit \\
\bottomrule
\end{tabular}
\end{table}

\section{Conclusion}
This work proposed the C-AoEI metric to quantify the fundamental freshness-reliability-efficiency trade-off in S-IoT systems constrained by propagation delay, dynamic fading, and bandwidth limitations. We designed a satellite-terrestrial architecture employing shadowed-Rician fading and an enhanced L-HARQ scheme with packet-level encoding, enabling parallel processing for multiple Ground Base Stations. A closed-form expression for C-AoEI was derived, explicitly characterizing the trade-off dynamics. Building on this, an adaptive optimization algorithm dynamically balances C-AoEI and transmission efficiency through joint adjustment of decision thresholds and coding parameters. 

\section*{Appendix: Relationship Analysis of \(\boldsymbol{\varepsilon_s(z)}\) with \(\boldsymbol{\omega_i}\) and \(\boldsymbol{\beta}\)}
This appendix provides a formal derivation of the mathematical relationships between the function \(\varepsilon_s(z)\) and the parameters \(\omega_i\) and \(\beta\), based on the governing equation:

\begin{equation}
\varepsilon_s(z) = \frac{ i - \dfrac{\ln \left( \omega_i \right)}{\beta} }{ S_z }.
\label{eq:main}
\end{equation}
where: \(\varepsilon_s(z)\) is the target function (e.g., strain or error), \(i\) is a constant index or fixed parameter, \(\omega_i > 0\) is a positive real variable (exponential weight), \(\beta \neq 0\) is a non-zero constant (typically \(\beta > 0\), decay rate), \(S_z > 0\) is a positive scaling factor (assumed independent of \(\omega_i\) and \(\beta\)).
\subsection*{1. Relationship between \(\boldsymbol{\varepsilon_s(z)}\) and \(\boldsymbol{\omega_i}\) (Fixed \(\boldsymbol{\beta}\))}

The equation is reformulated as:
\begin{equation}
\varepsilon_s(z) = \underbrace{\frac{i}{S_z}}_{A} - \underbrace{\frac{1}{S_z \beta}}_{B} \ln \left( \omega_i \right),
\label{eq:omega}
\end{equation}
where \(A = \dfrac{i}{S_z}\) and \(B = \dfrac{1}{S_z \beta}\) are constants. This reduces to:
\[
\varepsilon_s(z) = A - B \ln \left( \omega_i \right),
\]
\textbf{Functional Dependence}:  
  \(\varepsilon_s(z)\) exhibits a linear relationship with \(\ln \left( \omega_i \right)\), indicating logarithmic dependence on \(\omega_i\).

\textbf{Monotonicity}:  
   For \(B > 0\) (standard case where \(\beta > 0\) and \(S_z > 0\)):  
    \(\varepsilon_s(z)\) \textbf{monotonically decreases} with increasing \(\omega_i\).  
    \textit{Mathematical Justification}:  
    \[
    \frac{\partial \varepsilon_s}{\partial \omega_i} = -\frac{B}{\omega_i} < 0 \quad (\because B > 0, \omega_i > 0).
    \]
    
    For \(B < 0\) (non-standard case, e.g., \(\beta < 0\) or \(S_z < 0\)):  
    \(\varepsilon_s(z)\) \textbf{monotonically increases} with \(\omega_i\).
  
\textbf{Key Properties}:  
 \textbf{Functional Dependence}:  
  \(\varepsilon_s(z)\) exhibits a linear relationship with \(\ln \left( \omega_i \right)\), indicating logarithmic dependence on \(\omega_i\).
  
\textbf{Monotonicity}:  
    For \(B > 0\) (standard case where \(\beta > 0\) and \(S_z > 0\)):  
    \(\varepsilon_s(z)\) \textbf{monotonically decreases} with increasing \(\omega_i\).  
    \textit{Mathematical Justification}:  
    \[
    \frac{\partial \varepsilon_s}{\partial \omega_i} = -\frac{B}{\omega_i} < 0 \quad (\because B > 0, \omega_i > 0).
    \]
    
    For \(B < 0\) (non-standard case, e.g., \(\beta < 0\) or \(S_z < 0\)):  
    \(\varepsilon_s(z)\) \textbf{monotonically increases} with \(\omega_i\).
  
  At \(\omega_i = 1\):  
    \[
    \left. \varepsilon_s(z) \right|_{\omega_i=1} = A = \dfrac{i}{S_z},
    \]
    
 Sensitivity to \(\omega_i\):  
    \[
    \left| \frac{\partial \varepsilon_s}{\partial \omega_i} \right| = \frac{|B|}{\omega_i} \propto \omega_i^{-1}.
    \]
    Sensitivity decreases inversely with \(\omega_i\).

\subsection*{2. Relationship between \(\boldsymbol{\varepsilon_s(z)}\) and \(\boldsymbol{\beta}\) (Fixed \(\boldsymbol{\omega_i}\))}

The equation is reformulated as:
\begin{equation}
\varepsilon_s(z) = \underbrace{\frac{i}{S_z}}_{P} - \underbrace{\frac{\ln \left( \omega_i \right)}{S_z}}_{Q} \frac{1}{\beta},
\label{eq:beta}
\end{equation}
where \(P = \dfrac{i}{S_z}\) and \(Q = \dfrac{\ln \left( \omega_i \right)}{S_z}\) are constants. This reduces to:
\[
\varepsilon_s(z) = P - Q \cdot \beta^{-1}
\]

 \textbf{Functional Dependence}:  
  \(\varepsilon_s(z)\) is linear in \(\beta^{-1}\), indicating inverse dependence on \(\beta\).
  
\textbf{Monotonicity} (assuming \(S_z > 0\)):  
 If \(\omega_i > 1\) (\(\ln(\omega_i) > 0 \implies Q > 0\)):  
    \(\varepsilon_s(z)\) \textbf{monotonically increases} with \(\beta\).  
    \textit{Mathematical Justification}:  
    \[
    \frac{\partial \varepsilon_s}{\partial \beta} = \frac{Q}{\beta^2} > 0 \quad (\because Q > 0, \beta^2 > 0)
    \]
    
 If \(\omega_i < 1\) (\(\ln(\omega_i) < 0 \implies Q < 0\)):  
    \(\varepsilon_s(z)\) \textbf{monotonically decreases} with \(\beta\).  
    \textit{Mathematical Justification}:  
    \[
    \frac{\partial \varepsilon_s}{\partial \beta} = \frac{Q}{\beta^2} < 0 \quad (\because Q < 0)
    \]
    
   If \(\omega_i = 1\) (\(\ln(\omega_i) = 0 \implies Q = 0\)):  
    \(\varepsilon_s(z) = P\).

    Asymptotic limits:  
    \[
    \lim_{\beta \to \infty} \varepsilon_s(z) = P = \dfrac{i}{S_z}, \quad \lim_{\beta \to 0^+} \left| \varepsilon_s(z) \right| = \infty
    \]
    Sensitivity to \(\beta\):  
    \[
    \left| \frac{\partial \varepsilon_s}{\partial \beta} \right| = \frac{|Q|}{\beta^2} \propto \beta^{-2}
    \]
    Sensitivity decreases quadratically with \(\beta\).

\subsection*{Summary of Functional Dependencies}

\begin{table}[h]
\centering
\caption{Summary of relationship trends (\(S_z > 0\), \(\beta > 0\))}
\vspace{0.3cm}
\begin{tabular}{lccc}
\toprule
\makecell{Fixed\\Parameter} & Variable & \makecell{Functional\\Relationship} & Trend Direction \\
\midrule
\(\beta\) & \(\omega_i\) & \(\varepsilon_s \propto -\ln(\omega_i)\) & Decreases \\[2pt]
\(\omega_i\) & \(\beta\) & \(\varepsilon_s \propto \beta^{-1}\)&\makecell{Increases ($\omega_i > 1$)\\Decreases ($\omega_i < 1$)}\\
\bottomrule
\end{tabular}
\end{table}

\vspace{1em}
\noindent \textbf{Note on Non-Standard Cases}:  
The direction of monotonicity reverses under the following conditions:
\begin{itemize}
  \item \(S_z < 0\) (negative scaling factor)
  \item \(\beta < 0\) (negative decay rate)  
\end{itemize}
In such cases, re-evaluate the partial derivatives using the signed constants \(B\) and \(Q\).

\color{black}

\newpage

 \begin{IEEEbiography}[{\includegraphics[width=1in,height=1.25in,clip,keepaspectratio]{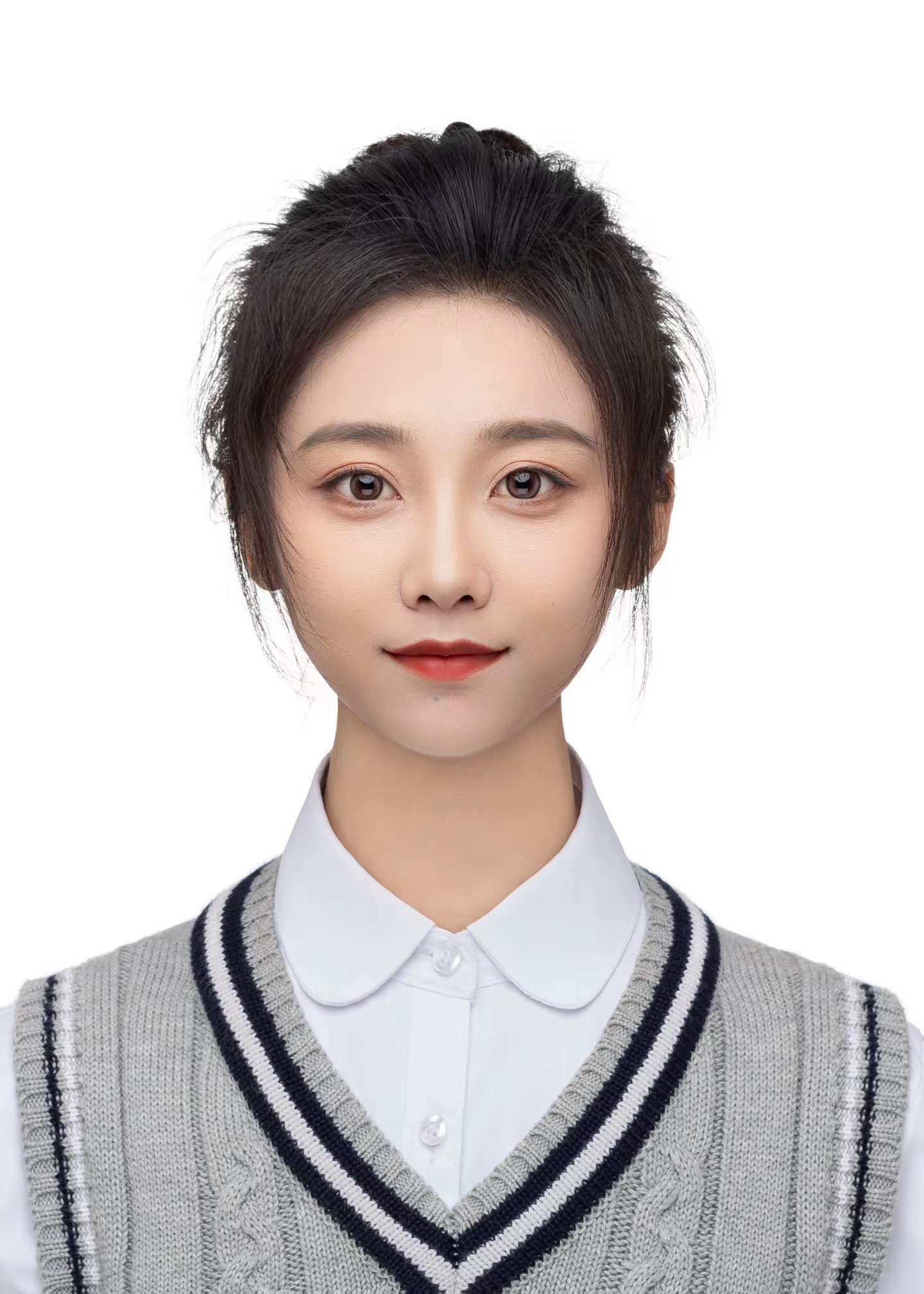}}]{Yuhua Zhao} received the B.S. and M.S. degrees in Major from Northwest Normal University (NWNU), Lanzhou, China, in 2020 and 2023, respectively.
She is currently pursuing the Ph.D. degree with the School of Information and Communication Engineering, Beijing University of Posts and Telecommunications (BUPT), Beijing, China. She has authored or coauthored several papers in refereed journals and conference proceedings. Her research interests include satellite communications, integrated satellite-terrestrial networks, and satellite navigation.
 
 \end{IEEEbiography}

 \begin{IEEEbiography}[{\includegraphics[width=1in,height=1.25in,clip,keepaspectratio]{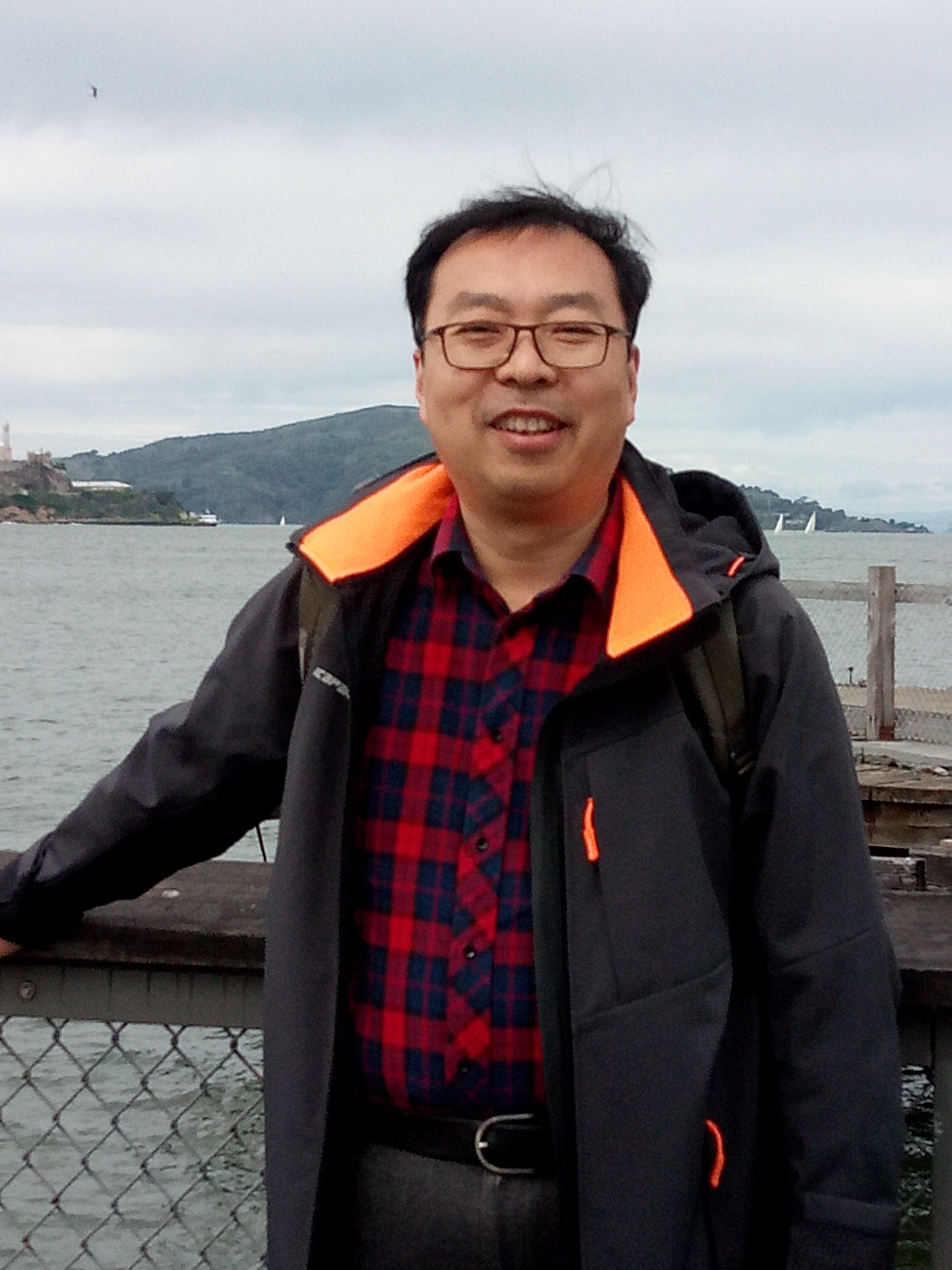}}]{Tiejun Lv} received the M.S. and Ph.D. degrees in electronic engineering from the University of Electronic Science and Technology of China (UESTC), Chengdu, China, in 1997 and 2000, respectively. From January 2001 to January 2003, he was a Post-Doctoral Fellow at Tsinghua University, Beijing, China. In 2005, he was promoted to a Full Professor at the School of Information and Communication Engineering, Beijing University of Posts and Telecommunications (BUPT). From September 2008 to March 2009, he was a Visiting Professor with the Department of Electrical Engineering, Stanford University, Stanford, CA, USA. He is currently the author of four books, one book chapter, and more than 160 published journal articles and 200 conference papers on the physical layer of wireless mobile communications. His current research interests include signal processing, communications theory, and networking. He was a recipient of the Program for New Century Excellent Talents in University Award from the Ministry of Education, China, in 2006. He received the Nature Science Award from the Ministry of Education of China for the hierarchical cooperative communication theory and technologies in 2015 and Shaanxi Higher Education Institutions Outstanding Scientific Research Achievement Award in 2025.
 
 \end{IEEEbiography}

 \begin{IEEEbiography}[{\includegraphics[width=1in,height=1.25in,clip,keepaspectratio]{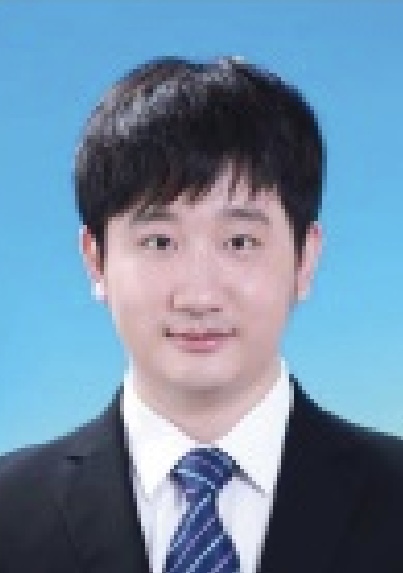}}]{Ke Wang} received his Ph.D degree in the College of Information and Communication of Beijing University of Posts and Telecommunications (BUPT), in 2014. He is now an associate professor in BUPT since 2019. He has published over 50 research papers in journals and conferences such as IEEE Transactions on Vehicular Technology and ICC, Globecom, etc. He is also served as the TPC member of several conference, such as ICC2015/2016/2017, WCNC 2019/2020/2021, PIMRC 2019/2020. He has hold 14 patents from the China Patent Office. He is the program leader of the National Key Research and Development Program of China. His research interests include integrated space-ground network technology, simulation technology and link-level signal processing.
 
 \end{IEEEbiography}



\vfill

\end{document}